\documentclass{aa}    
\usepackage{graphicx,url,twoopt,natbib}
\usepackage[varg]{txfonts}           
\usepackage{hyperref}					
\hypersetup{
  colorlinks=false,   
  urlcolor=blue,     
  linkcolor=red,     
}

\bibpunct{(}{)}{;}{a}{}{,}    

\makeatletter
\renewcommand*\aa@pageof{, page \thepage{} of \pageref*{LastPage}}
\makeatother

\begin{document}  

\titlerunning{Characterizing young substellar binaries}
\authorrunning{Calissendorff et al.}

\title{Spectral characterization of newly detected young substellar binaries with SINFONI}

\author{Per Calissendorff$^{1}$ \and Markus Janson$^1$\and Rub\'{e}n Asensio-Torres$^1$ \and Rainer K{\"o}hler$^{2,3}$ 
                                }

\institute{
        Department of Astronomy, Stockholm University, Stockholm, Sweden\\
        e-mail: {\bf per.calissendorff@astro.su.se}
\and
		 Sterrewacht Leiden, P.O. Box 9513, NL-2300 RA Leiden, The Netherlands
\and
		 University of Vienna, Department of Astrophysics, Türkenschanzstr. 17 (Sternwarte), A-1180 Vienna, Austria    
}        


\abstract{
We observe 14 young low-mass substellar objects using the VLT/SINFONI integral field spectrograph with laser guide star adaptive optics to detect and characterize 3 candidate binary systems. All 3 binary candidates show strong signs of youth, with 2 of them likely belonging to young moving groups. Together with the adopted young moving group ages we employ isochrones from the BT-Settle CIFIST substellar evolutionary models to estimate individual masses for the binary components. We find 2MASS J15104786-2818174 to be part of the $\approx 30 - 50$ Myr Argus moving group and composed of a $34 - 48\,M_{\rm Jup}$ primary brown dwarf with spectral type M$9\gamma$ and a fainter $15 - 22\, M_{\rm Jup}$ companion, separated by $\approx 100$ mas. 2MASS J22025794-5605087 is identified as an almost equal-mass binary in the AB Dor moving group, with a projected separation of $\approx 60$ mas. Both components share spectral type M$9\gamma/\beta$, which with the adopted age of $120 - 200$ Myr yields masses between $50 - 68\,M_{\rm Jup}$ for each component individually. The observations of 2MASS J15474719-2423493 are of lesser quality and we obtain no spectral characterization for the target, but resolve two components separated by $\approx 170$ mas which with the predicted young field age of $30 - 50$ Myr yields individual masses below $20\,M_{\rm Jup}$. Out of the 3 candidate binary systems, 2MASS J22025794-5605087 has unambiguous spectroscopic signs of being a bona-fide binary, while the other two will require second-epoch confirmation. The small projected separations between the binary components corresponds to physical separations of $\approx 4 - 7$ AU, allowing for astrometric monitoring of just a few years in order to generate constrained orbital fits and dynamical masses for the systems. In combination with their young ages, these binaries will prove to be excellent benchmarks for calibrating substellar evolutionary models down to a very low-mass regime.
}  

\maketitle



\section{Introduction} \label{sec:intro}
Multiplicity studies of stars and substellar objects have revealed evidence for a decreasing binary frequency as a function of spectral type. Substellar binaries are thus relatively rare compared to their more massive stellar counterparts, with more than $70\,\%$ of massive stars of B and A-type to be in binary systems or part of higher hierarchical systems \citep{Kouwenhoven+07, Peter+12}. Massive young stellar objects also seem to have higher binary fraction compared to T Tauri and main sequence O-type stars \citep{Pomohaci+19}, suggesting for age and evolutionary stage to affect multiplicity rates. For Solar-type stars the binary frequency decreases to $\approx 50\,\%$ \citep{DM91, Raghavan+10} and decreases even further to $\approx 40\,\%$ for M-stars \citep{FM92, Delfosse+04, Janson+12}. For later M-types, stretching down to early L-types and into the substellar regime, the binary fraction decreases to $\approx 15\,\%$ \citep{Close+03}. Due to the smaller sample sizes, the binary rate for the substellar population is much less constrained, with varying binary frequencies ranging $10 - 20\,\%$ for old brown dwarfs (BDs) of ages 1 - 10 Gyr in the field \citep{Bouy+03,  Burgasser+06}. Highlighting the trend of decreasing binary fraction with later spectra type, \citet{Fontanive+18} reports a binary frequency of $\approx 5\,\%$ for field BDs of spectral types T5-Y0 at separations of 1.5 - 1000 AU. Along with the trend of decreasing binary frequency with decreasing primary mass, which appears persistent throughout both the stellar and substellar mass-regimes, the semi-major axis distribution also decreases to smaller separations for smaller primary mass \citep{KH12,  DK13}. Additionally, the mass ratio distribution is shifted towards unity, with higher mass ratios $q \equiv M_2/M_1$ for lower primary mass binaries \citep{Burgasser+07}, implying the binary population in the substellar regime to be more compact and symmetric compared to earlier types.

Statistics for multiplicity in the substellar regime remain poorly constrained however, as large suitable samples to study have been lacking until recent years. Large sky-surveys like the 2 Micron All Sky Survey \citep[2MASS;][]{2MASS}, Wide-field Infrared Survey Explorer \citep[WISE;][]{Wright+10} and the Panoramic Survey Telescope And Rapid Response System\footnote{\url{http://pswww.ifa.hawaii.edu/pswww/}} (Pan-STARRS) have increased the number of known substellar objects substantially. As such, several substellar binaries in the Galactic field have been identified; many being old, cool, and of very late spectral type \citep[e.g.][]{Gelino+11, Dupuy+15, Huelamo+15}. Efforts have also been put into constraining young stellar associations and Young Moving Groups (YMGs) \citep[e.g.][]{Malo+13, Gagne+14}, enabling the possibility to identify young substellar objects and BD binaries such as 2MASS J1207334-393254 \citep{Chauvin+04} and 2MASS J11193254-1137466 \citep{Best+17}. Trends found in multiplicity studies over both a younger and an older population such as frequencies, semi-major axis distribution and mass-ratio distributions can thus be used to inform theories about the formation and evolution of brown dwarfs \citep[e.g.][]{MC14}. However, some formation scenarios for brown dwarfs such as the ejection scenario \citep{RC01}, turbulent fragmentation \citep{PN04} and binary disruption \citep{GW07} only predict for tight binaries on separations smaller than $\approx 10$ AU to survive to old field ages, suggesting that wide BD binaries such as WISE J1217+1626 and WISE J1711+3500 \citep{Liu+12} may form in an alternative formation channel. 

Brown dwarfs with constrained ages have also lately become important in the exoplanet aspect, as low-mass BDs have many characteristics analogous to directly imaged exoplanets such as atmospheres and some formation scenarios. Most detected BD binaries have been of nearby sources \citep[e.g. Luhman 16 AB, $\approx 2$ pc;][]{Luhman13}, with many being old and cool BDs in the field \citep[e.g. WISE J014656.66+423410.0;][]{Dupuy+15}. As YMGs and associations are being more well-defined, we are starting to discover more young BD binaries, including those within the planetary-mass regime \citep[e.g. 2MASS J11193254-1137466;][]{Best+17}. Finding these low-mass BDs in YMGs and on tight orbits allows for constraining dynamical masses of the systems within just a few years of monitoring, which can be compared to and calibrate evolutionary models. The importance of obtaining suitable benchmark substellar objects has been well-advocated, as discrepancies are systematically seen when comparing luminosity-derived masses from models to dynamical masses \citep[e.g.][]{Dupuy+14}. Inconsistencies are also frequently visible when comparing isochrone ages of substellar and low-mass companions to earlier type stellar primaries \citep[e.g.][]{CJ18, Asensio-Torres+19}.

In this paper we present a search for companions to young BDs at the very bottom end of the substellar mass-regime, near the classical Deuterium burning limit at around $M = 13 - 17\,M_{\rm Jup}$ \citep[e.g.][]{Spiegel+11}, in order to identify potential touchstone systems to test against formation and evolutionary models. The small sample of objects observed in the study makes it difficult to constrain the multiplicity frequency for young low-mass BDs, but we discover some systems of interest for which multiplicity studies may be expanded upon. We describe the observations including the sample selection and reduction in Section~\ref{sec:obs}. We perform a spectral analysis of our targets and present our findings in Section~\ref{sec:res}. A more detailed description of the observed targets is shown in Section~\ref{sec:binaries} for the binaries and in Section~\ref{sec:singles} for the unresolved and single sources, with astrometric solution of the detected binaries tabulated in Table~\ref{tab:results}. We summarise our results and discuss them in Section~\ref{sec:disc}.

\section{Method} \label{sec:method}

\subsection{Sample selection}
The observed sample consists of 14 sources out of a larger list of 22 low-mass brown dwarfs obtained from the BANYAN survey \citep{Gagne+15}, where we put constraints on the targets in terms of estimated mass and age. We also include the additional constraint of only choosing sources that have nearby bright enough tip-tilt stars so that they can be observed with laser guide assisted adaptive optics (AO). The targets are identified as low-mass objects with some indications of youth; either by having low surface gravity, or high membership probability of being part of a YMG. We estimate from the YMGs that the targets have ages ranging 12- 200 Myr, depending on specific group membership, and are all located within 70 pc of the Sun. Full names of the observed targets and YMG membership status are given in Table~\ref{tab:obs}, where names are otherwise abbreviated from the full 2MASS identifier to the form 2Mhhmm.

\subsection{Observations} \label{sec:obs}
The observations were based on the European Southern Observatory 0101.C-0237(A) program and carried out through May to September in 2018 in Service mode at the Very Large Telescope (VLT) at Cerro Paranal, Chile. The instrument used for the observations was the Spectrograph for INtegral Field Observations in the Near Infrared \citep[SINFONI;][]{Bonnet+04, Eisenhauer+03}, where all observations were carried out with the $H+K$ grating and 1500 spectral resolution power. The adaptive optics  is required in order to resolve companions on small separations to our targets. Due to the faint nature of the targets, the observations employed the Laser Guide Star (LGS) facility that was available for SINFONI at the time of the observations (SINFONI has since been removed from the telescope and placed on another unit telescope without LGS). Tip-tilt stars were obtained for all targets in order to optimise the LGS-AO observations. Each target was observed with an exposure time of 300 seconds that was repeated 6 times with a small dithering pattern used to correct for sky and bad pixels. The procedure of the data reduction is described in Section~\ref{sec:red}.

The program observed 14 targets in total. However, due to difficulties in acquisition for such faint sources under modest or poor conditions, only 8 of the observations showed clear signs of the target sources, with 6 of them being of sufficiently good quality that spectra could be extracted from the data. For one of the 2 bad cases, 2MASS J15474719-2423493 (hereafter 2M1547), we find that the source is a tentative planetary-mass binary that will require follow-up observations to confirm. In the other poor quality case, 2MASS J18082129-0459401 (hereafter 2M1808), the source is only seen in 2 of the 6 frames, with the object being very close to the edge of the detector so that potential companions outside would go undetected. We hence exclude 2M1808 completely from our analysis. For the 6 high quality sources we find 2 to be good substellar binary candidates. The observations and the status of the sources are listed in Table~\ref{tab:obs}, also showing the YMGs affiliated with the respective sources.

\begin{table*}[t]
\centering
{
\caption{Log of VLT/SINFONI observations.}
\begin{tabular}{lccccc}
\hline \hline
 Object ID & Short & Obs. Date & YMG & Membership prob. & Status\\
\hline
2MASS J02103857-3015313 & 2M0210 & 2018 Aug 22 & THA & 99.9 & No source detected \\ 
2MASS J14252798-3650229 & 2M1425 & 2018 May 23 & ABD & 99.9 & No source detected \\
2MASS J15104786-2818174 & 2M1510 & 2018 Jul 15 & ARG & 99.2 & Binary candidate \\ 
2MASS J15474719-2423493 & 2M1547 & 2018 May 23 & Field & -  & Binary candidate \\ 
2MASS J17125805+0645416 & 2M1712 & 2018 May 23 & - & - & No source detected \\
WISE J18082129-0459401$^{\rm a}$ & 2M1808 & 2018 Jul 15 & ARG & 94.5 & Poor quality \\ 
2MASS J20113196-5048112 & 2M2011  & 2018 May 29 & THA & 42.6 & No source detected \\
2MASS J20135152-2806020 & 2M2013 & 2018 Jun 06 & BPMG & 99.7 & No source detected \\
2MASS J20334473-5635338 & 2M2033 & 2018 Aug 18 & THA & 93.4 & Unresolved source \\ 
2MASS J21171431-2940034 & 2M2117 & 2018 Aug 06 & BPMG & 98.4 & Unresolved source \\ 
2MASS J22025794-5605087$^{\rm b}$ & 2M2202 & 2018 Aug 17 & THA-ABD & 98.4 - 32.1 & Binary candidate \\ 
2MASS J22351658-3844154 & 2M22351 & 2018 Sep 27 & THA & 96.2 & No source detected \\
2MASS J22353560-5906306 & 2M22353 & 2018 Aug 09 & THA & 99.9 & Unresolved source \\ 
2MASS J23225299-6151275 & 2M2322 & 2018 Sep 18 & THA & 96.7 & Unresolved source \\ 

 \\
\hline
\label{tab:obs}
\end{tabular}

\raggedright {\footnotesize {\bf Notes:} Young Moving Group and associations are Argus (ARG), AB Doradus moving group (ABD), Tucana-Horologium (THA), $\beta$ pic Moving Group (BPMG). Membership probabilities are obtained from the BANYAN $\Sigma$ online tool.

$^{\rm a}:$ Membership probabilities are based on the BANYAN $\Sigma$-tool using optimal parallax and radial velocity for Argus group. With no prior on those quantities the membership probability is reduced to 3.4 \%.

$^{\rm b}:$ Different young moving groups for different version of the BANYAN tool. The former value depicts the result from the BANYAN II tool and paper that discovered the object, whereas the latter value shows the updated group values for the BANYAN $\Sigma$-tool.}

}
\end{table*}

\subsection{Data reduction} \label{sec:red}
\begin{figure}[hbtp]
  \centering
  \includegraphics[width=\columnwidth]{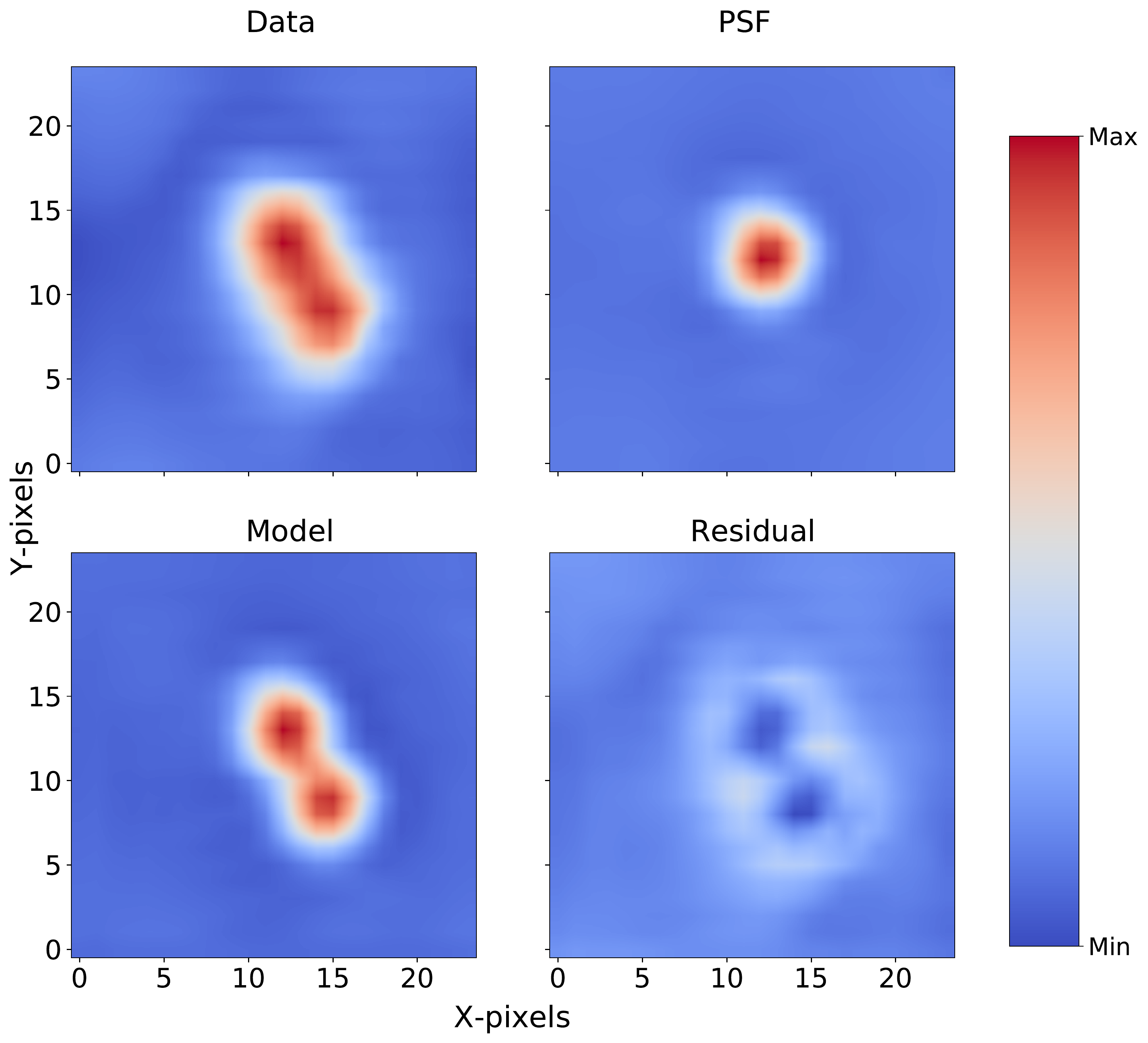}
  \caption[]{\label{fig:psf_match} %
 Plots of the PSF-matching for the binary 2M2202. The upper left and right plots shows the observed data and the PSF of the standard star used respectively. The lower left plot show the best-fit of a constructed model made from 2 reference PSF that we match to the positions and scale with the brightness of the 2 components. The lower right plot depicts the residual after subtracting the model from the observed data. All plots have been normalised to match the observed data.
   }
\end{figure}

Raw data is reduced using the ESO reduction pipeline, {\it Esoreflex} \citep{Esoreflex}, where raw images are converted to full data cubes. We note that there is a drift in \emph{x} and \emph{y} positions for different wavelengths within the cube, where the peak value of the sources changes from the shortest wavelength slice to the longest wavelength, likely caused by atmospheric dispersion. We solve the issue with the drift by applying a development-level version of the pipeline which corrects for the atmospheric dispersion. The full data cubes are then collapsed and investigated by eye to identify potential companions. For detected potential binary candidates we employ an iterative PSF fitting method to account for partially overlapping PSFs of the closely separated binaries. The procedure is similar to the scheme developed for the AstraLux M-dwarf multiplicity survey \citep{Janson+12, Janson+14}, also described in \citet{Calissendorff+17}. Concisely, the scheme takes a rough initial estimation for the location in $x$ and $y$ pixels along with a prediction of the brightnesses of the 2 components in the binary. Subsequently, we build a model system using 2 copies of a PSF reference, typically the standard stars observed close in time to the observations of the target. The brightnesses and positions of the 2 components are then varied iteratively, and subtracted from the observed data until a minimum residual solution is found. The procedure is illustrated in Figure~\ref{fig:psf_match} for the target 2M2202 which serves as an example. The usage of several different standard stars as PSF provides us with some statistical variations of the astrometric positions, letting us evaluate errors resulting from imperfections from the PSF matching. As the PSF fitting procedure yields differential coordinates $\Delta x$ and $\Delta y$ in pixels, we convert the output into sky coordinates using the plate scale from the SINFONI user manual\footnote{\url{http://www.eso.org/sci/facilities/paranal/instruments/sinfoni/doc/VLT-MAN-ESO-14700-3517_v93.2.pdf}} of $p_x = 12.5$ mas and $p_y = 25$ mas. The PSF fitting procedure is performed on each individual wavelength-slice within the data cube. We vary the PSF reference for each binary candidate and keep the median astrometric value of each wavelength slice, with the standard deviation as uncertainty. We then again take the median of the astrometry for the entire wavelength range as our final result, presented in Table~\ref{tab:results}, where we combine the statistical errors from the median wavelength range and the uncertainty in individual wavelength slices.

\begin{figure*}[hbtp]
  \centering
  \includegraphics[width=1.3\columnwidth]{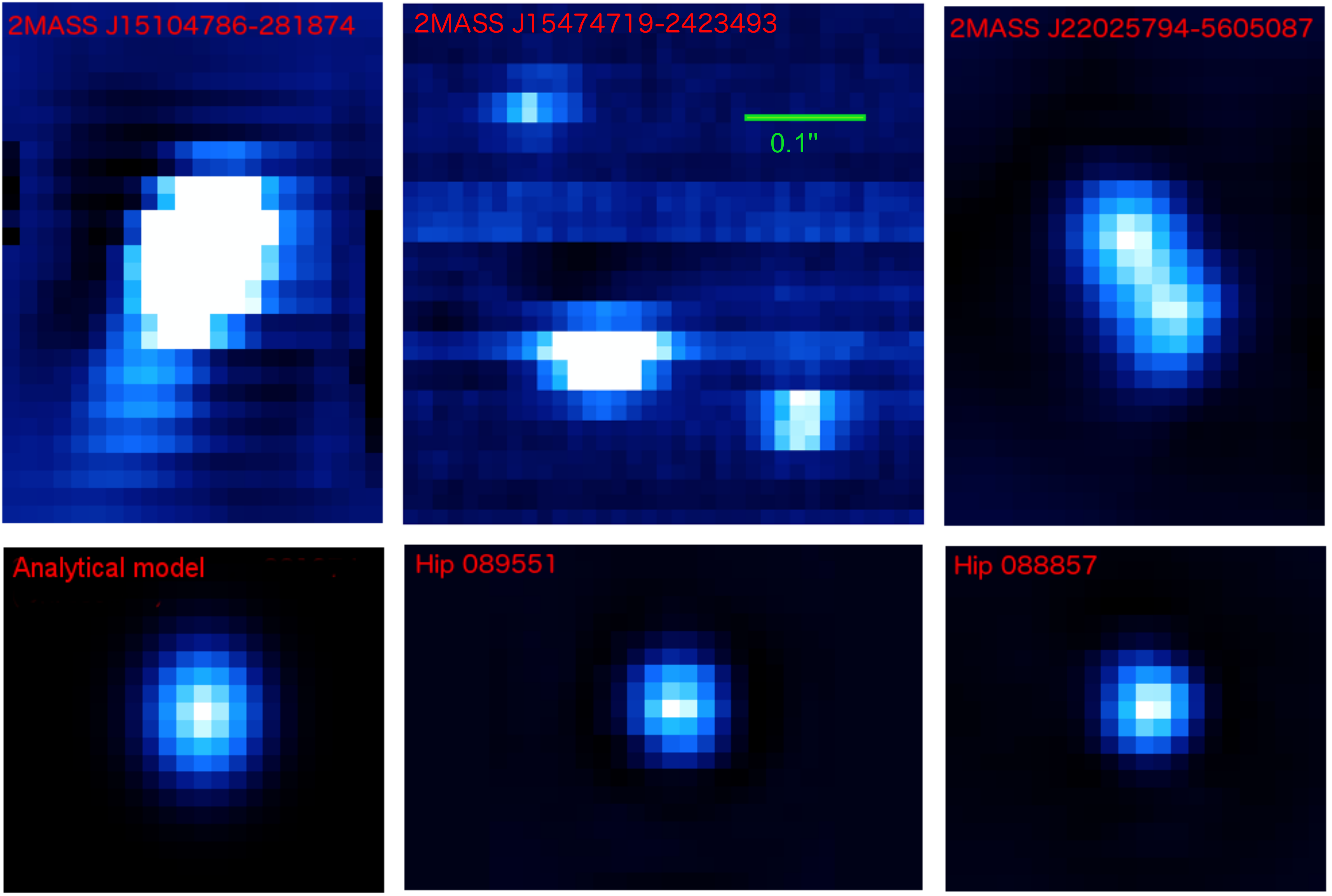}
  \caption[]{\label{fig:binaries} %
  Upper images show the collapsed data cubes of all observed frames for the detected binary candidates; 2M1510 (left), 2M1547 (middle), and 2M2022 (right). The lower images depict the corresponding relevant point source reference used in the fitting procedure of the astrometric measurements. Special care is taken for 2M1510 where we find no suitable standard reference star to match the PSF of the brown dwarf binary components, and thus an unresolved 2D Gaussian function is fitted to the entire system which is further explained in Section~\ref{sec:2M1510}. For 2M1547 we note a third point source in the upper left part of the image which we do not include in our analysis since it was weak and appeared artefact-like in our original data reduction that did not correct for atmospheric dispersion. Derived astrometric parameters and photometry for each individual object are seen in Table~\ref{tab:results}. The scaling in each individual image is chosen to better highlight the individual components. North is up and East is to the left in the figures.
   }
\end{figure*}

All targets are handled in the same way with the exception of 2M1547, where the observations are of poor quality and the source appears much fainter than expected, hence we obtain an ambiguous detection of the source. The observations of 2M1547 also lack proper sky subtraction due to the small dithering pattern employed, causing self-subtraction of the source signal when attempting to coadd frames. Thus, we reduce the frames with the pipeline using no in-built sky subtraction and perform custom post-subtraction using empty sky frames obtained directly before and after the observations of 2M1547. The resulting collapsed data cubes for each detected binary candidate are shown in Figure~\ref{fig:binaries} together with their respective point spread calibrator reference.

\section{Results} \label{sec:res}

\subsection{Colour-magnitude diagrams} \label{sec:phot}
The PSF matching algorithm employed also provides us with information of the relative brightness of each component in the matching, which is then used to calculate the magnitude of the individual components. We choose a subset of wavelength-slices within the data cubes which we use to construct a custom $H'$ magnitude between wavelengths of $1.5 - 1.8\,\mu$m and a custom $K'$ magnitude with wavelengths ranging $2.1 - 2.45\,\mu$m. We calculate the apparent magnitude in $H'$ and $K'$ band separately by using the unresolved apparent magnitudes from the 2MASS catalogue \citep{2MASS} of each system and scaling the brightness with respect to the relative brightness of the components acquired from the PSF matching, then adding up the total flux for each band and compare it to literature 2MASS fluxes.  The PSF matching routine is performed several times, varying the reference PSF in order to obtain a statistical measurement for the relative brightnesses. We produce a median relative brightness for each individual wavelength-slice, and keep the standard deviation as the uncertainty, which is then propagated together with the formal 2MASS unresolved magnitude errors to estimate the uncertainty in our derived magnitudes. The resulting apparent magnitudes are shown in Table~\ref{tab:results}, where we use GAIA DR2 parallaxes to calculate distances and absolute magnitudes.

\begin{figure}[hbtp]
  \centering
  \includegraphics[width=\columnwidth]{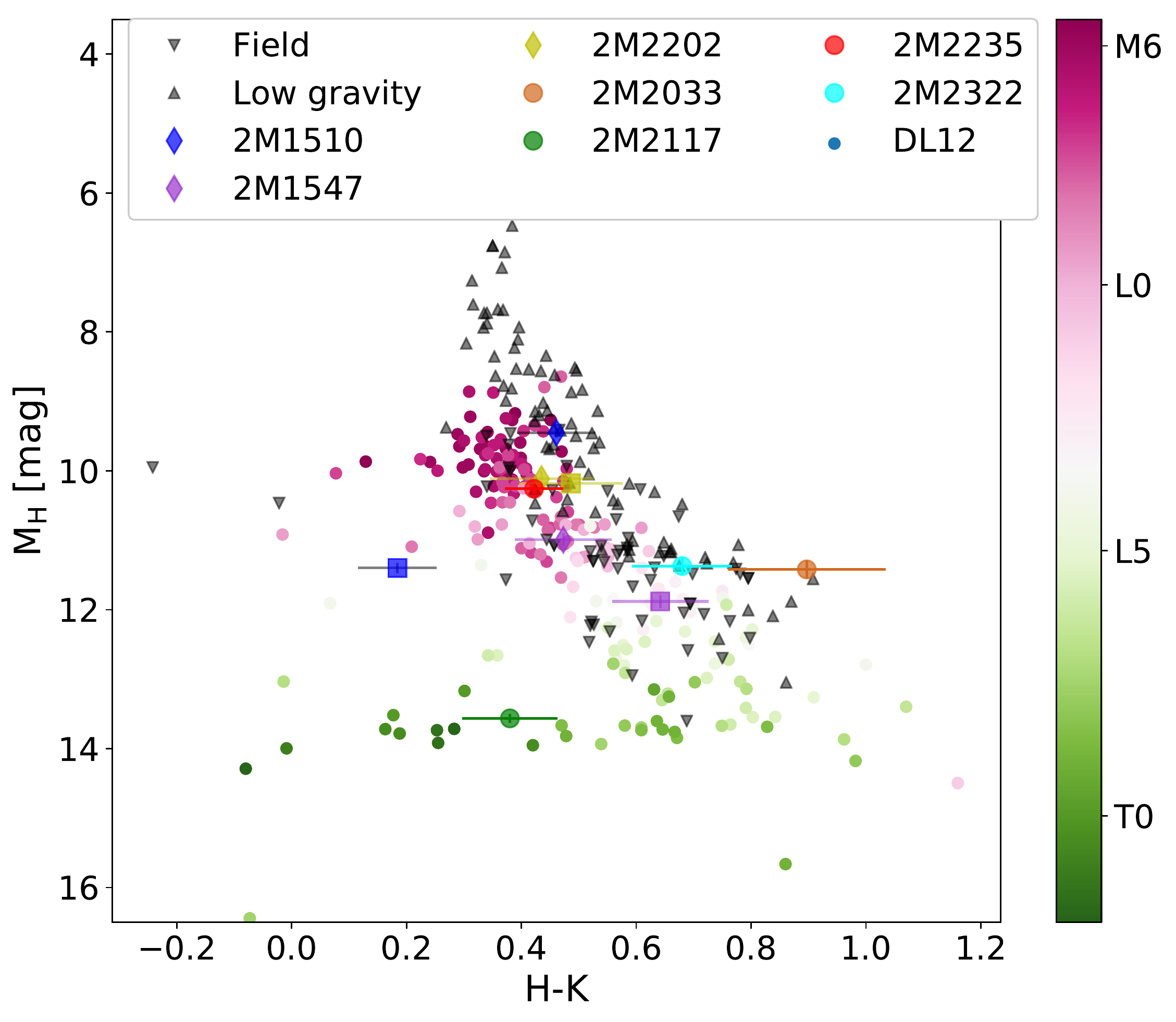}
  \caption[]{\label{fig:CMD} %
	Colour-magnitude diagram showing our observed targets together with known substellar objects from the literature. The blue, magenta and yellow markers show the binary candidates 2M1510, 2M1547 and 2M2202 respectively, where the primaries are illustrated by the diamonds and the squares represent the secondary binary components. Unresolved and single sources in our survey are shown by circles only. We include low-mass stars and BDs from \citet{DL12} as smaller circles, which have been colour coded on a spectrum shown by the colour bar in the figure, going from late M type to early T type. We only include objects from the list that have known spectral type and Gaia parallaxes. The gray triangles show low-mass objects from the BANYAN survey \citep{Gagne+15}, where the downward pointing triangles show field dwarfs, and the upward pointing triangles objects with low surface gravity and strong signs of youth. Photometry is based on 2MASS bands for all sources. Most of our sources seems to be on the redder side in the diagram, suggesting for some extinction by being dusty and can be considered as a sign of youth. 
   }
\end{figure}

Colour-magnitude diagrams (CMD) can reveal some insight to youth and spectral types of these kind of substellar objects. In Figure~\ref{fig:CMD} we place our targets in a CMD and compare them to similar objects found in the literature, including low-mass stars and known BDs from \citet{DL12} as well as BDs with low gravity and field BDs from the BANYAN survey \citep{Gagne+15}. Most of our observed targets, both single and binaries, show signs of low surface gravity and considered young as seen in the diagram, with the exception of 2M2117 which is a later spectral type ($\sim$T0) compared to the other objects in our target sample, where most are between M9-L1 spectral types.

\begin{table*}[t]
{
\caption{Compiled results for the observed targets including photometry and astrometry for the binary candidates.}
\begin{tabular}{lcccccccc}
\hline
 Object & SpT type & Distance & $H'_{\rm app}$ & $K'_{\rm app}$  & Age & Mass & Separation & P.A. \\ 
 & & [pc]  & [mag]  & [mag] &  [Myr] & [$M_{\rm Jup}$] & [mas] & [deg] \\ 
\hline
2M2033 & L$0\gamma \pm 1$ & $55.4 \pm 3.6 ^{\bf a}$ & $15.14 \pm 0.11$ & $14.24 \pm 0.08$ & $20 - 40$ &  $15.72^{+2.54}_{-3.56}$  & & \\
2M2117 & T$0$ & $16.9 \pm 1.7^{\rm \bf b}$ & $14.53 \pm 0.04$ & $14.15 \pm 0.07$ &  $21 - 26$ & $8.07^{+0.66}_{-0.98}$ & & \\
2M22353 & M$9\gamma \pm 1$ & $46.42 \pm 0.72$ & $13.59 \pm 0.04$ & $13.17 \pm 0.03$ &  $ 20 - 40$ & $20.62^{+3.50}_{-5.83}$ & & \\
2M2322 & L$1\gamma \pm 2$ & $43.03 \pm 1.94 $ & $14.54 \pm 0.06$ & $13.86 \pm 0.04$ &  $20 - 40$ &  $15.93^{+2.35}_{-3.57}$ & &  \\ \\

\hline
2M1510A & M$9\gamma_{-1}$  & $ 36.74 \pm 0.36$ & $12.28 \pm\, 0.04$& $11.82 \pm\, 0.04$& $30 - 50$ & $41.50^{+7.04}_{-7.37}$  & & \\ 
2M1510B &  L$1\beta$(?)  & -  & $14.22 \pm\, 0.06$& $14.09 \pm\, 0.04$& $30 - 50$ & $17.68^{+4.20}_{-2.10}$ & $103.43 \pm 13.43 $ & $64.96 \pm 2.12$ \\ 
2M1547A &  & $34.25 \pm 0.46$  & $13.67 \pm\, 0.07$& $13.19 \pm\, 0.04$& $30 - 50$ & $19.51^{+4.48}_{-2.30}$  & & \\ 
2M1547B &   & - & $14.55 \pm\, 0.07$& $13.91 \pm\, 0.04$& $30 - 50$ & $15.51^{+3.85}_{-2.05}$ & $ 173.66 \pm 0.59 $ & $15.62 \pm 0.50$ \\ 
2M2202A & M$9\gamma/\beta \pm 1$ & $69.97 \pm 2.36$ & $14.34 \pm\, 0.07$ & $13.91 \pm\, 0.06$ & $120 - 200$ & $57.37^{+11.30}_{-7.90}$ & & \\ 
2M2202B & M$9\gamma/\beta \pm 1$  & - & $14.41 \pm\, 0.08$ & $13.92 \pm\, 0.06$ & $120 - 200$ & $55.60^{+10.89}_{-7.43}$ & $ 60.36 \pm 2.67 $ & $ 59.51 \pm 1.58$\\ 
 \\
\hline
\label{tab:results}
\end{tabular}
\raggedright {\footnotesize {\bf Notes:}
 Masses are calculated from BT-Settl evolutionary models. 

$^{\rm \bf a}:$ Distance measurement from \citet{Gagne+15}.
$^{\rm \bf b}:$ Distance measurement from \citet{Best+15}.
}
}
\end{table*}


\subsection{Spectral analysis}\label{sec:spec}
We obtain unresolved spectra for each of our targets by fitting circular apertures to the target source, where the radii of the apertures are determined by first matching a 2D Gaussian to the wavelength-collapsed cube, separately for the $H'$ and $K'$ bands. The same is done for the standard stars obtained for each target, which is then used to flux-calibrate the source spectra. In order to obtain resolved spectra for the detected binaries we scale the unresolved spectra for each component by their relative brightness obtained from the PSF-matching described in Section~\ref{sec:red}, where the standard deviation for the relative brightness in each wavelength slice is adopted as the uncertainty for each respective spectral element. The resolving power of SINFONI with the $H+K$-band is $R = 1500$, corresponding to a spectral resolution of $\Delta \lambda = 5\,\AA$. 

To match model spectra with the observed flux from our target sources we adopt the same approach as \citet{Cushing+08}, and scale the models to match the observations and estimate a goodness-of-fit value for each model. The data is then compared to the modelled spectral energy distribution, where the goodness-of-fit statistic $G_k$ for each model $k$ is defined by

$$\label{eq:GoF}
G_k = \sum^n_{i=1} \left( \frac{f_i - C_k F_{k,i}}{\sigma_i} \right)^2,
$$

with $n$ being the number of data pixels, $i$ the index of the wavelength, $f_i$ the observed flux, $F_i$ the modelled flux, $\sigma_i$ the uncertainty and $C_k$ the scaling constant used. We calculate the scaling constant $C_k$ for each spectral band $H'$ and $K'$ separately as

$$
C_k = \frac{\sum f_i F_{k,i}/\sigma^2_i}{\sum F^2_{k,i}/\sigma^2_i},
$$

and select the best fitting models by locating the minimum $G$-values for 3 cases of $H'$-band only, $K'$-band only and $H'+K'$ bands together, ignoring the wavelength-range between the bands at around $\approx 1.8 - 2.1\,\mu$m. Because the wavelength dispersion is constant in both bands for SINFONI, we ignore the weight of the wavelength used in the original equation. We perform the spectral analysis for each target and component by matching the observed spectra with empirical templates to obtain the spectral type, as well as comparing the observed spectra to theoretical spectra from the BT-Settl CIFIST \citep{Baraffe+15} evolutionary models, described further in the following sections. 

\subsubsection{Empirical templates}
\begin{figure}[hbtp]
  \centering
  \includegraphics[width=\columnwidth]{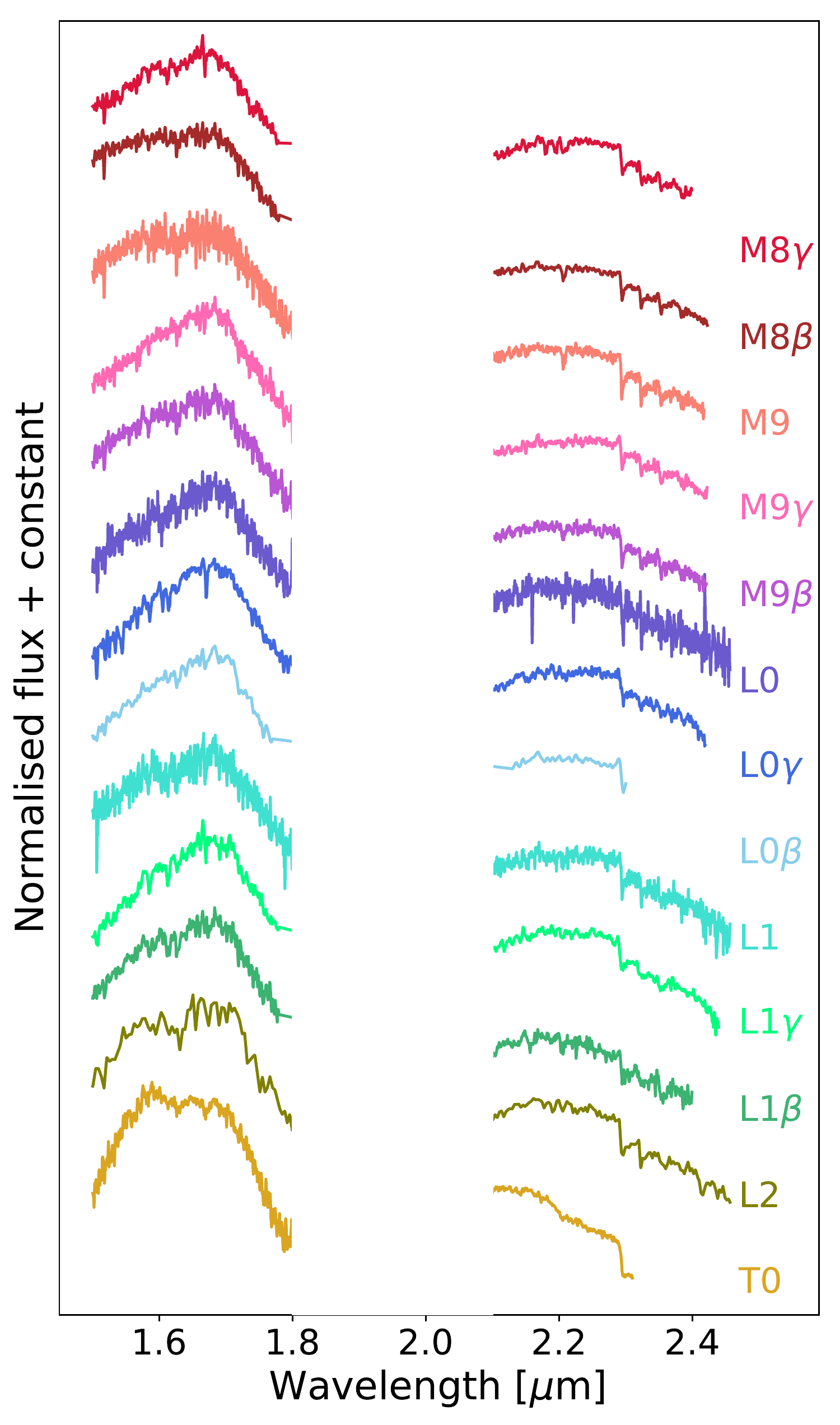}
  \caption[]{\label{fig:composite_templates} %
  Examples of the composite spectral templates produced from the median of several empirical spectra from the literature. The composite templates for late M and early spectral types with surface gravity indices $\gamma$ and $\beta$ are constructed by taking the median of several objects from the Montreal spectral library \citep{Gagne+15, Robert+16}. For T-types we construct median template spectra from known T-types in the SpeX Prism Library \citep{SpeX} and the Sloan Digital Sky Survey \citep{Chiu+06}. The templates have typical resolution of $\Delta \lambda \approx 5 - 40\,\AA$.
   }
\end{figure}

We create empirical templates of spectral types by combining and taking the median of several already known BD spectra from the literature. We adopt the same spectral indices as introduced by \citet{Kirkpatrick+06} that signify low surface-gravity and youth (i.e. $\alpha \sim 1000$ Myr, $\beta \sim 100 $ Myr, $\gamma \sim 10$ Myr, $\delta \sim 1$ Myr). We gather most of the empirical objects from the Montreal spectral library\footnote{\url{https://jgagneastro.wordpress.com/the-montreal-spectral-library/}} \citep{Gagne+15, Robert+16}, and produce our composite templates from the median of the sources available for a given spectral type and surface-gravity index, selecting only the empirical sources which share the same spectral resolution for a given spectral type. The empirical sources are first interpolated to have matching wavelengths before taking the median values to construct the composite templates, with resolutions varying between $\Delta \lambda \approx 5 - 40\,\AA$ from best to worst. The composite spectral templates are shown in Figure~\ref{fig:composite_templates}. For the later spectral types such as T-type dwarfs we do not procure surface-gravity indices, and instead form our median template model spectra from known T-types from the SpeX Prism Library \citep{SpeX} and the Sloan Digital Sky Survey \citep{Chiu+06}. 

\begin{figure}[hbtp]
  \centering
  \includegraphics[width=\columnwidth]{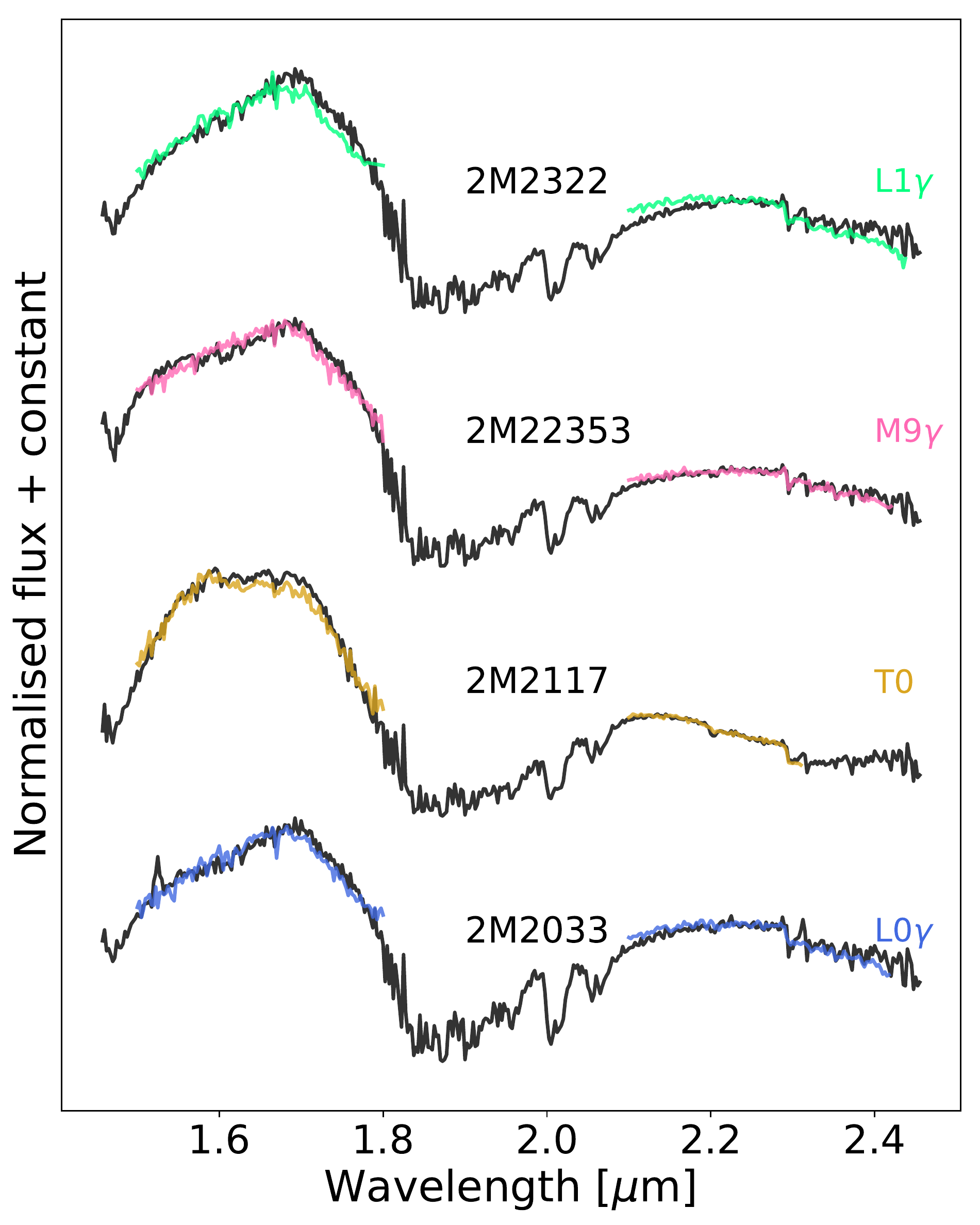}
  \caption[]{\label{fig:singlespecs} %
  Empirical spectral templates fitted to the unresolved single sources with no detected companions. The spectra are smoothed to match the resolution of the templates, and then scaled and fitted according to Equation~\ref{eq:GoF}.
   }
\end{figure}

Spectra from our observed sources are smoothed to match the resolution of the templates and subsequently interpolated to match the template wavelengths. For the objects identified as unresolved sources or single BDs in our survey we find good matches to their previously indicated spectral types from the BANYAN survey \citep{Gagne+15}. We show the best-fit templates to the observed spectra in Figure~\ref{fig:singlespecs}, where the fluxes have been normalised and scaled with a constant to fit the plot. For the binary candidates we were only able to obtain sufficiently good spectra for 2M1510 and 2M2202 to make a spectral analysis. We use same spectral type model templates as with the other sources, but scale the observed flux for the binaries according to their relative brightness. Due to the very small relative separations between the binary components in both cases, it is likely that the spectra of the companions will contain contaminations from their respective primary component. For 2M2202 this contribution is likely to be small, as the relative brightness is close to unity and we expect the spectral types to be the same. However, for 2M1510 the contamination is more prominent, and we obtain some degeneracies when matching the spectral type for the fainter companion. We estimate the uncertainty of the empirical fits as the range of templates with a goodness-of-fit within $\sqrt{2}$ of the best-fit value, typically corresponding $\pm 1$ spectral type for a given source.

\subsubsection{Theoretical models}
We compare the observed spectra to the theoretical BT-Settl CIFIST models from \citet{Baraffe+15}, where we convolve the spectral models to the width of the observed spectra and interpolate the wavelength steps to match the observations. For the theoretical model fits of the single or unresolved sources, we bin the observed data beforehand to a wavelength resolution of $\Delta \lambda = 25\,\AA$, adding the standard deviation of the mean of the bins to the uncertainty. We make use of the models that assume solar metallicity and have temperatures ranging $1200 - 2900$ K and surface gravities between $\log g = 3.5-5.0$, matching the theoretical spectra to the observations and estimate a goodness-of-fit value for each model. We note that there are several degeneracies within the models, where alternating between temperatures and surface gravities sometimes yield similar goodness-of-fit values. We plot the resulting best-fit models together with the observed spectra for each single-system source together with a grid-map showing the goodness-of-fit for each model in Figures~\ref{fig:2M2033} -~\ref{fig:2M2322}. The black dot in the grid-maps represents the best-fit models, and the contours depict the $68\%$, $95\%$ and $99\%$ confidence intervals calculated as $\Delta G_k = 2.3$, $6.18$ and $11.83$ respectively.

\subsection{Chance projections} \label{sec:projection}
\begin{figure}[hbtp]
  \centering
  \includegraphics[width=\columnwidth]{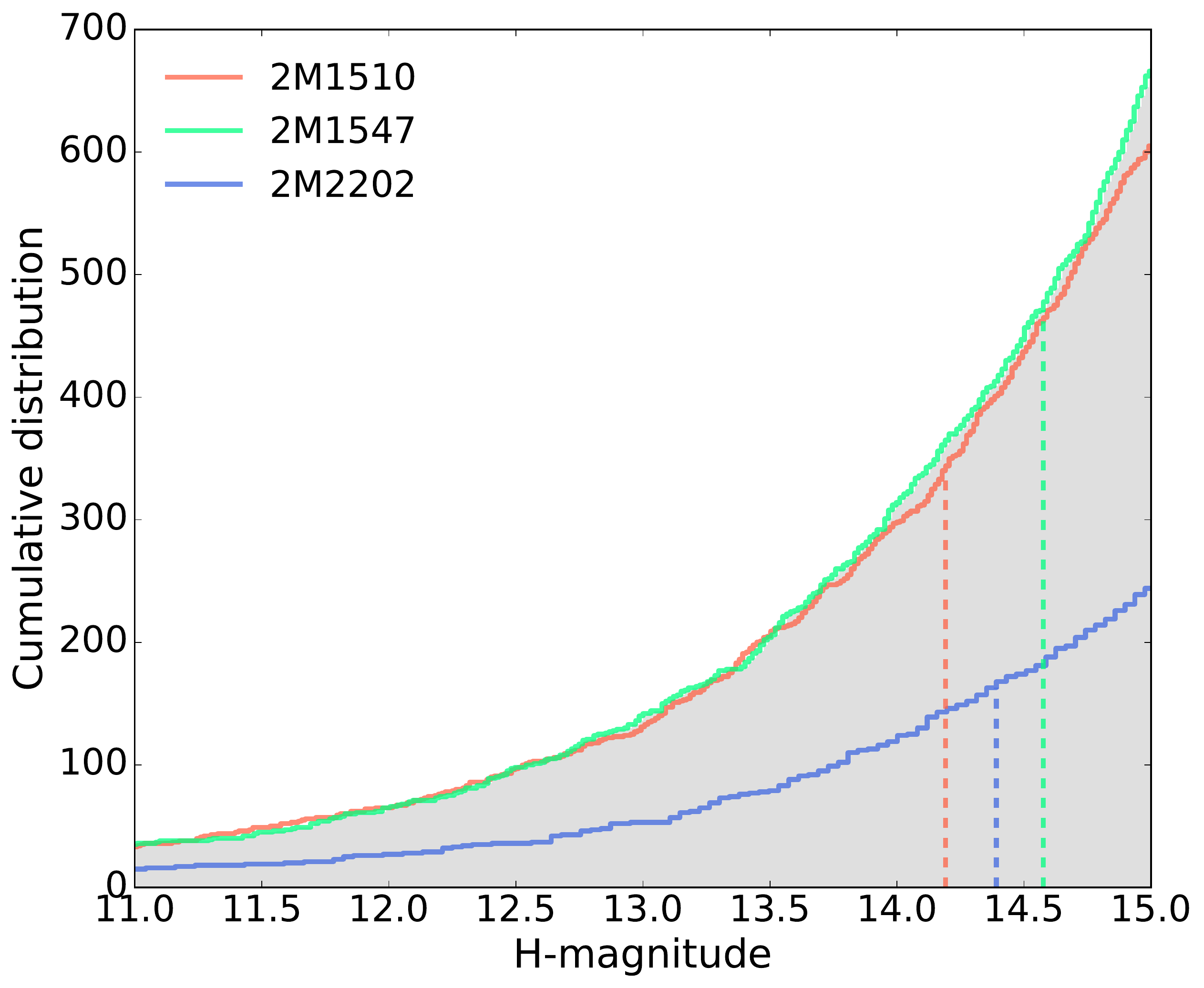}
  \caption[]{\label{fig:background} %
	Cumulative distribution of 2MASS point sources of different $H$-band magnitudes in a 15' radius around our detected binary candidates, represented by the red line for 2M1510, the green line for 2M1547 and the blue line for 2M2202. The dashed lines indicate the secondary component $H'$-band magnitude for each respective binary and the corresponding cumulative number of point sources that are brighter in the field. The field around 2M2202 is somewhat less dense than the other 2, and the estimated chance of having any of our binary candidates being a background source is on the order of $\sim 10^{-5}$ or smaller.
   }
\end{figure}

Considering the very small field of view employed by SINFONI of 0.8''$\times$0.8'', it is already unlikely that our detected binaries would be contaminated by background sources. However, in order to make a statistical argument about the binary components being gravitationally bound we calculate the probability of chance projection for background sources to our systems, applying the same procedure as described in \citet{Correia+06} and \citet{Lafreniere+14}. For this purpose we make use of the 2MASS Point Source Catalog\footnote{\url{https://irsa.ipac.caltech.edu/cgi-bin/Gator/nph-dd}} (PSC) and probe a larger field with a radius of 15' around each of our targets, and calculate a cumulative distribution of sources brighter in $H'$-band magnitude than the companion binary for each target, illustrated in Figure~\ref{fig:background}. By assuming that unrelated objects across the field are uniformly randomly distributed we obtain the average surface density of brighter objects, $\Sigma\, ({H} < {H'}_{\rm companion})$. The probability of having at least one unrelated source within a specific angular separation, $\Theta$, from a target of interest is then estimated by:

$$
P(\Sigma, \Theta) = 1 - e^{-\pi \Sigma \Theta^2}
$$

Applying the calculations for each of our binary system yield chance projections with probabilities of $1.2 \cdot 10^{-5}$ for 2M1510, $5.7 \cdot 10^{-5}$ for 2M1547 and $2.4 \cdot 10^{-6}$ for 2M2202. By assuming that all systems share the same source density as the most crowded field, and that all companions share the same brightness and separation as the faintest and most distant companion candidate observed, we can set an upper limit to the probability of observing 3 chance projections among the 7 target sources successfully observed. For a given source the estimated number of background sources is equivalent to the chance projections, which we estimate as $3.8 \times 10^{-4}$. Thus, assuming a Poisson distribution, the probability of obtaining 3 chance projections for all 7 of the observed sources is near negligible, $9.2 \times 10^{-12}$.

\subsection{Binary candidates} \label{sec:binaries}
\begin{figure}[hbtp]
  \centering
  \includegraphics[width=\columnwidth]{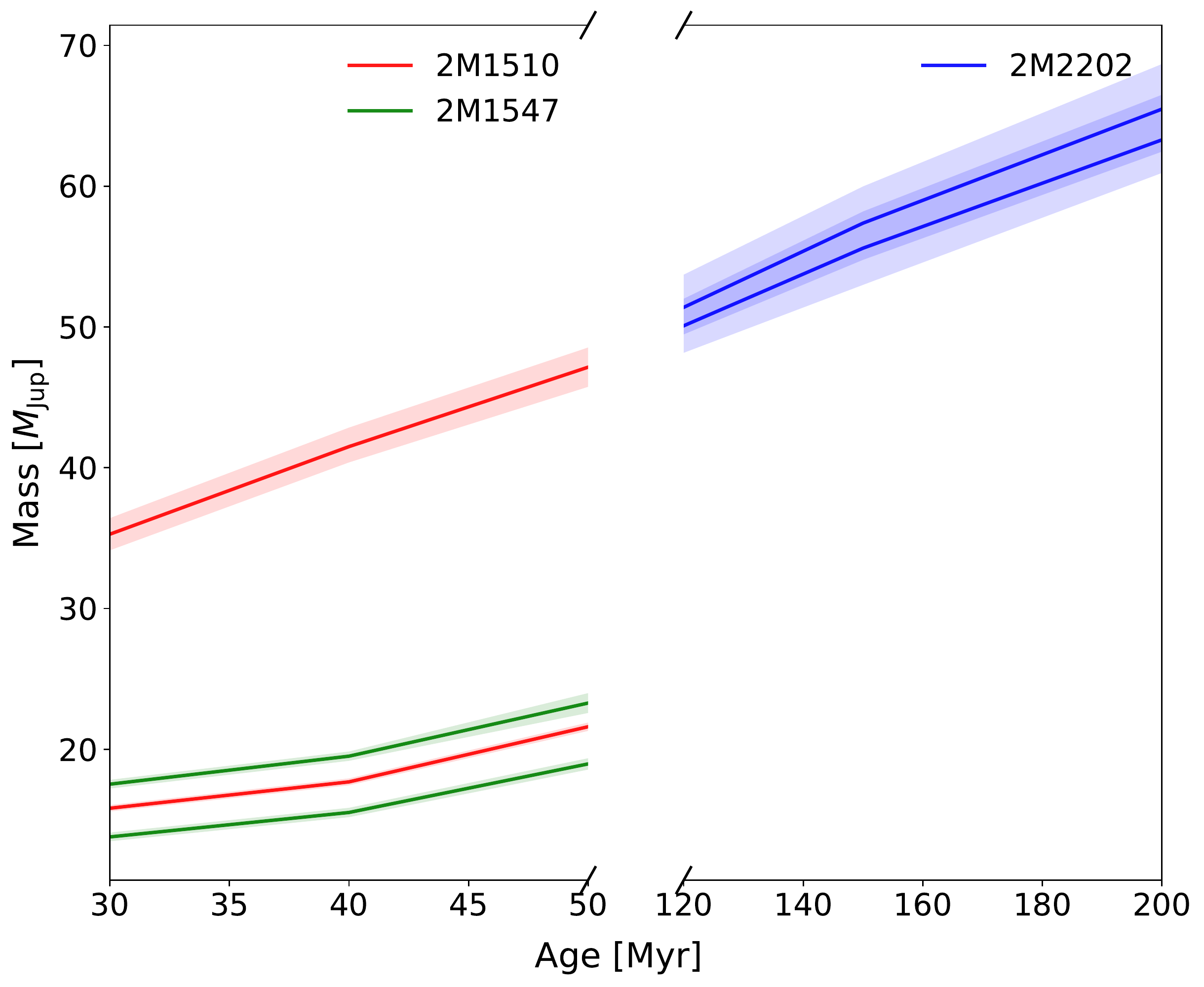}
  \caption[]{\label{fig:masses} %
  Isochrones showing the theoretical masses inferred from the BT-Settl CIFIST evolutionary models for the binaries presented in this work. The primary and secondary components are depicted by the upper and lower solid lines for each respective source, and the shaded coloured areas represents the uncertainty in the mass-estimates from the luminosities. The plot has a discontinuity in the age-range on the x-axis, showing only the ages of relevance for the objects included, hence not featuring ages between $50 - 120$ Myr. We note that alternative isochrone models may predict other masses for the same ages and luminosities, and we base our choice of models for consistency with the theoretical spectral models employed in our analysis.
   }
\end{figure}

By examining the wavelength-collapsed data cubes from our observations we detect 3 of our targets to be binary candidates, where we are also able to obtain resolved spectra for 2 of the targets. Assuming that all detected binaries are bona fide systems we can also assume the binary components to have the same age as their respective primary. By using the resolved brightness of the systems we can thus estimate individual masses by comparing their luminosity to isochrones from theoretical evolutionary models. Here we employ the BT Settl CIFIST 2011-2015 models from \citet{Baraffe+15} to be consistent with our spectral analysis. The astrometric data for the resolved binaries and inferred masses for the respective components along their corresponding ages and magnitudes are shown in Table~\ref{tab:results}, which are plotted in an age-mass diagram in Figure~\ref{fig:masses}. We combine the errors from our photometry with the age-ranges of the relevant YMGs to obtain the uncertainties in the mass estimated from the isochrones. 

The following sections provide further detailed information on the individual systems classified as binaries in our survey.

\subsubsection{2MASS J15104786-2818174} \label{sec:2M1510}
We identify 2M1510 as a substellar binary from the wavelength collapsed data cube in the leftmost image in Figure~\ref{fig:binaries}. The system has previously been identified as a late M-dwarf of spectral type M9 and a potential member of the Argus moving group with an age of $30 - 50$ Myr \citep{Burgasser+15, Gagne+15, Faherty+16}. We note that the Convergence Tool\footnote{\url{http://dr-rodriguez.github.io/CPCalc.html}} developed by \citet{Rodriguez+13} suggests the system to be in the $\sim 200$ Myr old Carina-Near YMG \citep{Zuckerman+06} with a $100\,\%$ probability. However, the Argus group is not featured in the Converge Tool, whereas the BANYAN $\Sigma$ includes both groups, and suggests a $99.2\,\%$ membership probability for Argus while dismissing Carina-Near completely. We assume the system to be in the Argus group, adopting an age of $30 - 50$ Myr for our subsequent calculations.

Gaia DR2 data yield proper motions of $\mu_{\rm RA} = -118.75 \pm 0.49$ mas/yr and $\mu_{\rm DEC} = -46.87 \pm 0.42$ mas/yr along with a system parallax of $27.22 \pm 0.27$ mas, giving 2M1510 a distance of about 37 pc \citep{Lindegren+18}. The projected separation between the primary and binary component is observed to be $103.43 \pm 13.43$ mas, translating to a physical separation of $\approx 4$ AU. The uncertainty in the astrometry for 2M1510 is greater than for the other binary candidates due to the lack of good enough PSFs to reproduce the data, and that most of the uncertainty stems from the $K'-$band in which the companion is fainter in comparison to at shorter wavelengths. In order to transform the projected separation to a statistical estimate for the semi-major axis we apply the conversation factor of 1.16 for very low-mass visual binaries derived by \citet{DL11}. We obtain a statistical estimate of the semi-major axis of $a \approx 4.4$ AU, which with Kepler's Third Law converts to an orbital period of $P \approx 30$ years. 

\begin{figure*}[hbtp]
  \centering
  \includegraphics[width=1.5\columnwidth]{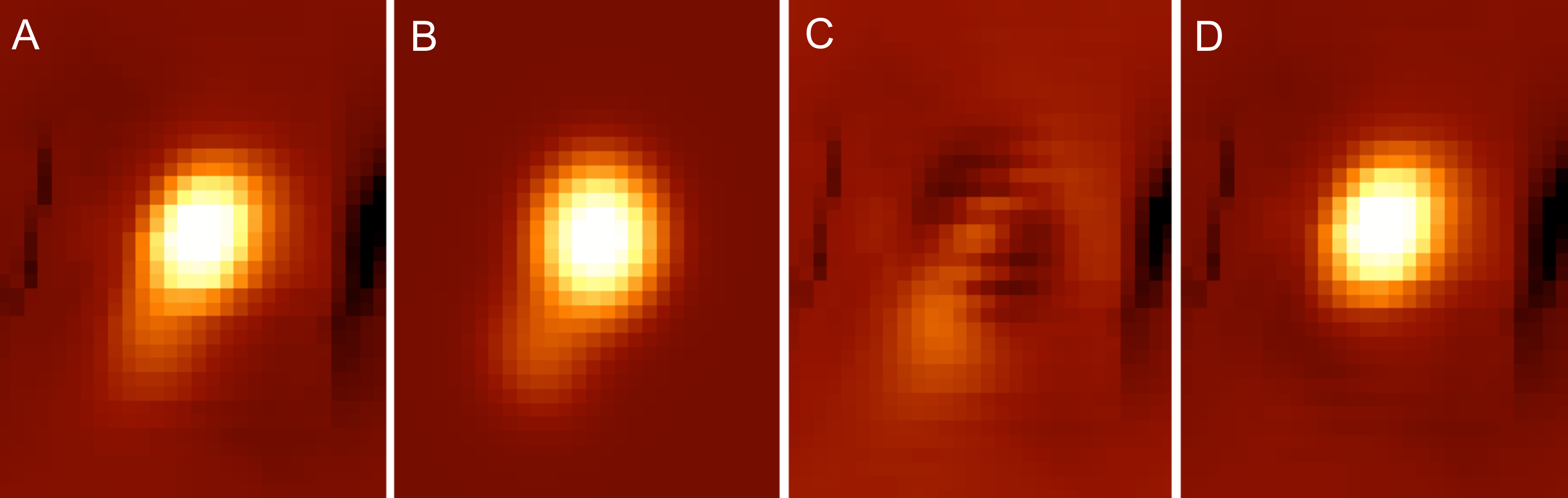}
  \caption[]{\label{fig:2M1510_res} %
 Snapshots of the collapsed data cube in the $H'+K'$-band for the binary candidate 2M1510. {\bf (A)} Observed data. {\bf (B)} Model of the data where the PSF has been placed over the primary and secondary positions and scaled to match their relative brightnesses. {\bf (C)} Residual after subtracting the model primary component from the observed data. {\bf (D)} Residual after subtracting the model secondary component from the observed data. All snapshots have been normalised to match the observed data, using a different scaling compared to Figure~\ref{fig:binaries}. The fainter component is visible in all of the data cubes obtained from the observations.
 }
\end{figure*}

We investigate whether the binary candidate companion to 2M1510 could be due to the effect of the PSF by plotting the residual left after subtracting the matched PSF model of only 1 component at a time, as illustrated in Figure~\ref{fig:2M1510_res}. The model is based on the unresolved system convolved with a 2D Gaussian fit, later scaled to match the brightness of each component and repositioned for the smallest residual as explained in Section~\ref{sec:red}. When subtracting the PSF model from the primary, the secondary shows a point source morphology. Additionally, no similar PSF effect is seen in the reference star observed just after the exposures of 2M1510. Furthermore, the astrometric behaviour is not consistent with a diffractive feature; while there is a $\sim 10\,\%$ difference in separation between the $H'$ and $K'$ band for astrometric measurements in the individual bands, it is much smaller than the $\sim 40\,\%$ effect that would be expected for diffractive features. We also scrutinise each separate data cube from the observations, verifying that the fainter source is real and visible in all cubes, and not due to a telescope pointing error where the primary could have been in the wrong position for a small fraction of the integration time. 

The observed resolved brightness of both components along with predicted masses are shown in Table~\ref{tab:results}. We estimate the individual masses by comparing the observed luminosities to isochrones from the BT-Settl CIFIST models, interpolating the model luminosities to the observed values and adopting the age of the YMG associated with the system. Doing so yields a primary mass of $M_{\rm primary} =  41.50^{+7.04}_{-7.37}\,M_{\rm Jup}$ and secondary mass ranging $M_{\rm secondary} = 17.68^{+4.20}_{-2.10}\,M_{\rm Jup}$, corresponding to a primary to secondary mass-ratio of $q \approx 0.45$.

\begin{figure*}[hbtp]
  \centering
  \includegraphics[width=\columnwidth]{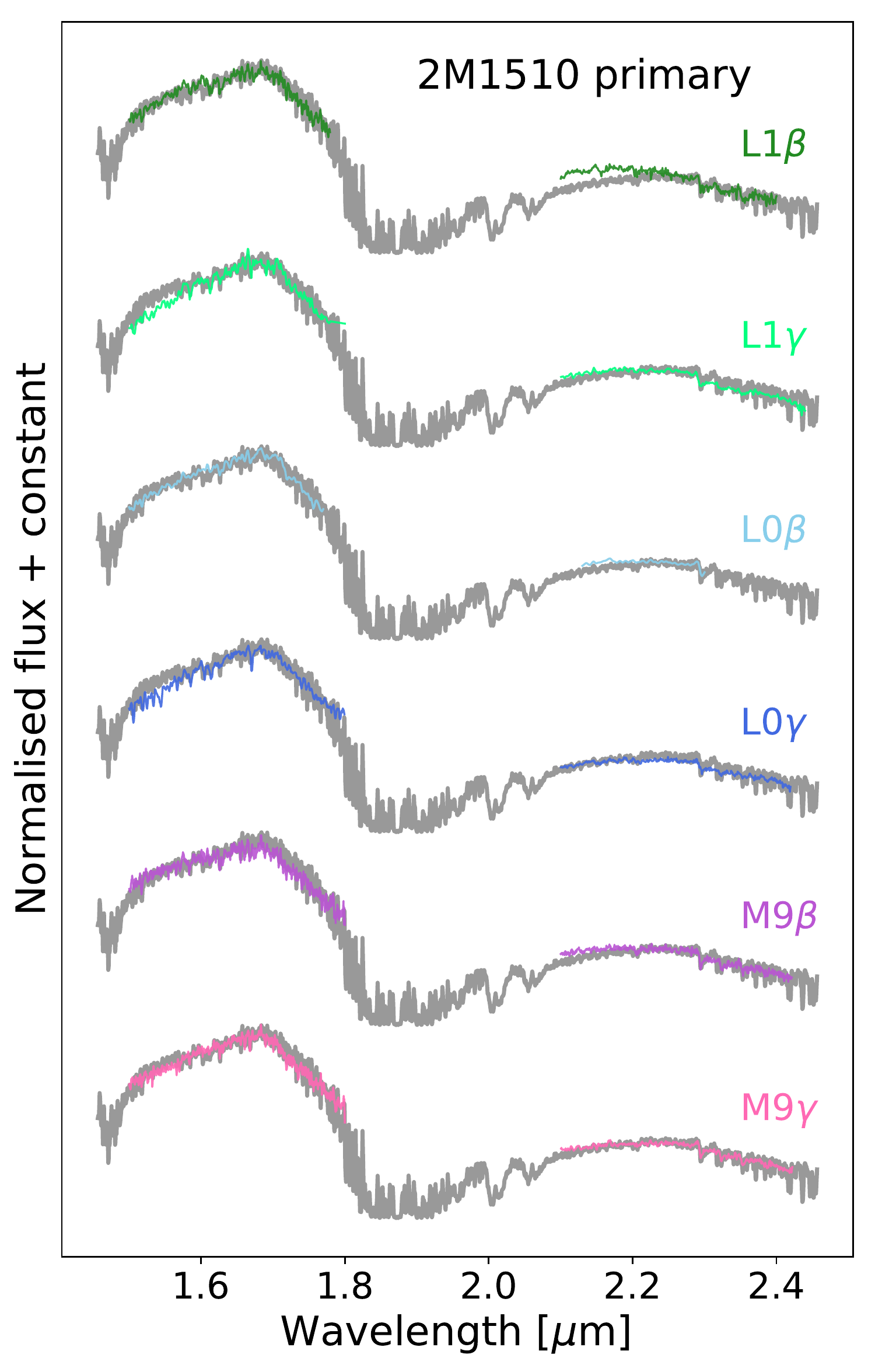}
  \includegraphics[width=\columnwidth]{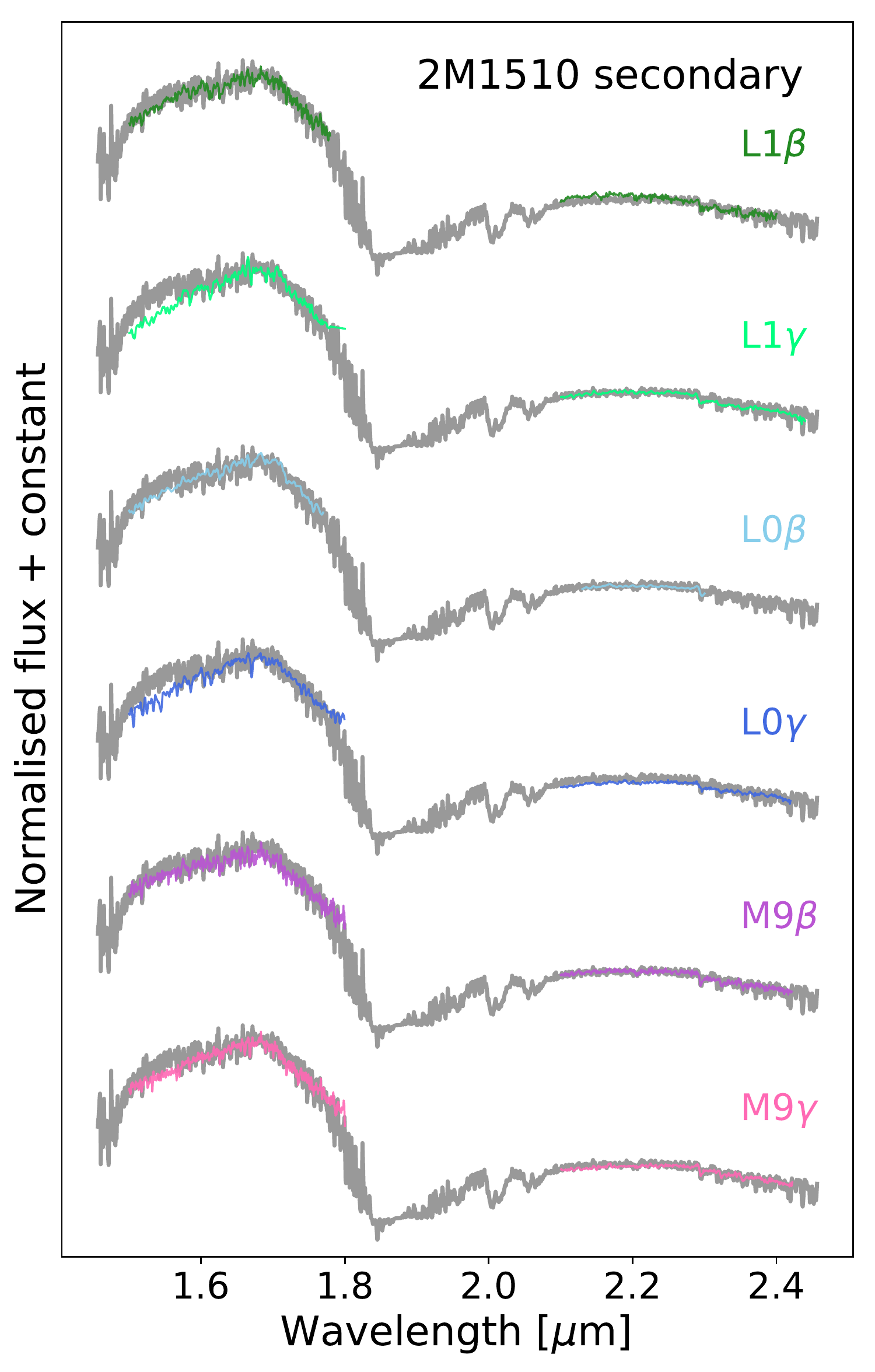}
  \includegraphics[width=\columnwidth]{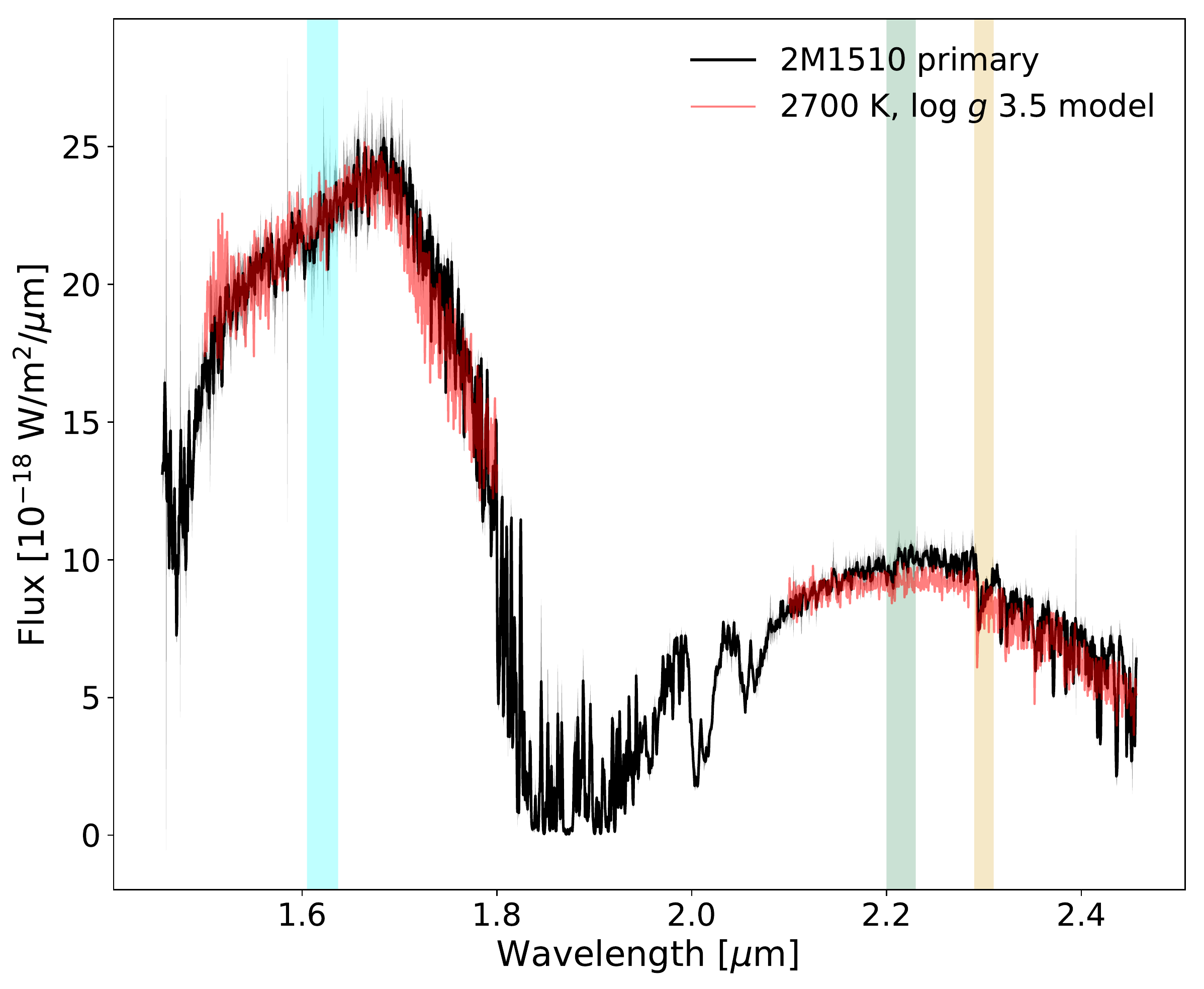}
   \includegraphics[width=\columnwidth]{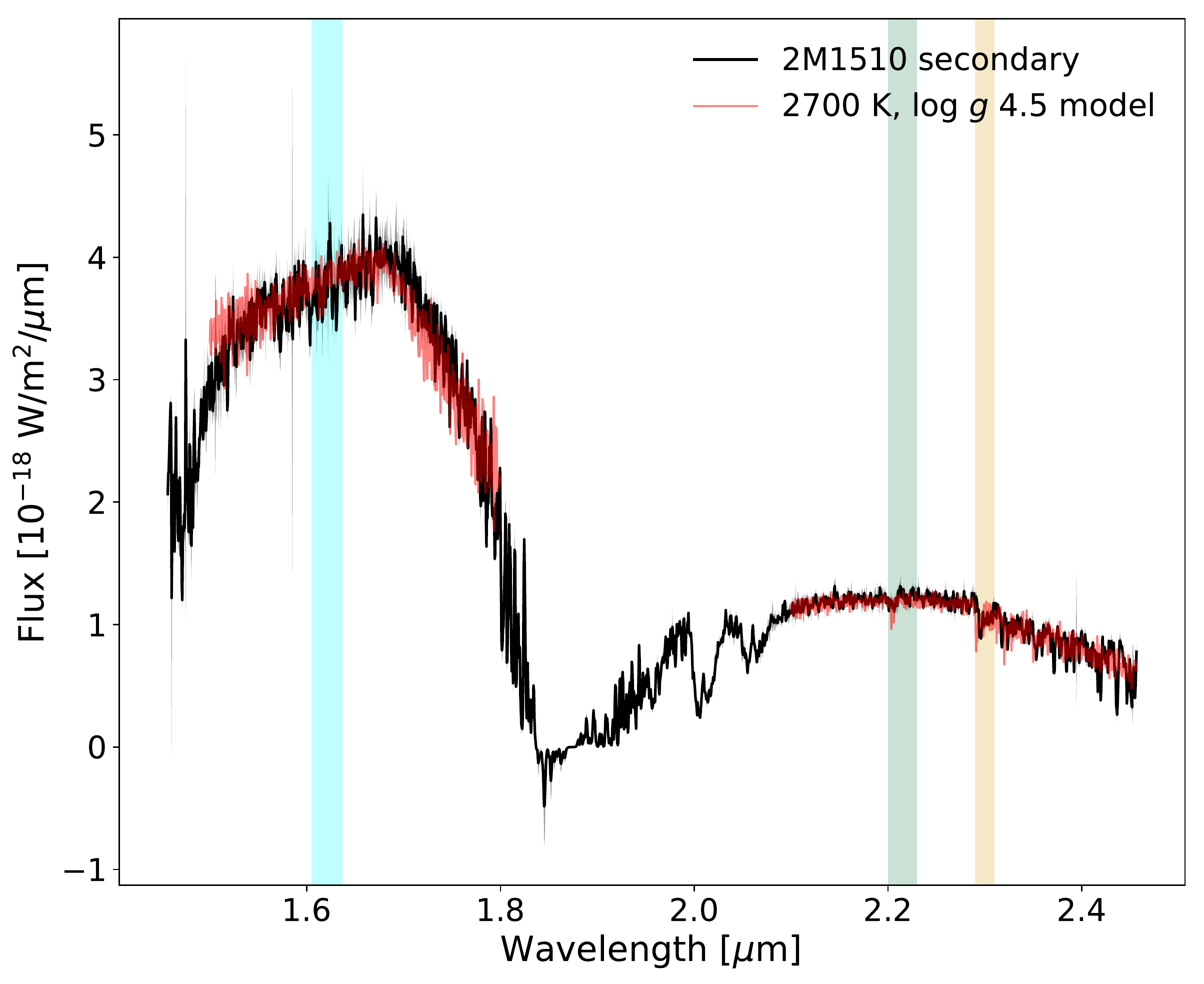}
  \caption[]{\label{fig:2M1510} %
Top row depicts empirical templates fitted to the spectra of the components to 2M1510, suggesting best-fits for $M9\gamma$ or earlier type for both components. Due to the contrast and proximity of the companion to the primary, it is very likely that much of the shape of the spectrum of secondary component is contributed by the primary. The companion is about a factor of 0.2 as bright as the primary, and it is unlikely they would be of the same spectral type (although this discrepancy may be mitigated if the primary is an unresolved binary itself). The lower plots show the best-fit theoretical evolutionary models from BT-Settl CIFIST to each individual component. We find the best-fit models to be $T_{\rm eff} = 2700$ K with $\log g = 3.5$ for the primary and $T_{\rm eff} = 2700$ K with $\log g = 4.5$ for the secondary. The contrast in magnitude between the components is estimated to be $\Delta {M_{\rm H}} \approx 2$, and we would expect the secondary component to be cooler than what the best-fit model we find predicts. 
   }
\end{figure*}

We compare the spectra of the binary pair according to the spectral analysis described earlier in Section~\ref{sec:spec}, using both empirical templates and theoretical models, as displayed in Figure~\ref{fig:2M1510}. From the spectral analysis we conclude that the primary, denoted as 2M1510A, is most likely a young M9$\gamma$ substellar object. The secondary, denoted as 2M1510B, proves more difficult due to the contamination from the primary and we are unable to verify a precise spectral type assignment. Nevertheless, the fainter nature of the companion and the flatter shape in the $K'$-band implies that it is of later spectral type than the primary. The theoretical models suggests for the primary to have an effective temperature of $T_{\rm eff} = 2600 \pm 100$ K and a surface gravity between $\log g = 4.0 \pm 0.5$. The best fit models for the companion candidate are very similar to the primary, with $T_{\rm eff} = 2600 - 2700$ K and $\log g = 4.0 - 5.0$, but much of the shape of the spectra is highly contaminated by the primary as seen in the lower plots of Figure~\ref{fig:2M1510} where we compare the 2 components side to side. The theoretical spectra for the primary are consistent with the effective temperature obtained from the isochrones for the same brightness. For the secondary component the theoretical spectra predict a much higher temperature than the isochrones, which would place the temperature $ T_{\rm eff} \leq 1800$ K. We believe this discrepancy is likely to stem from the high contamination from the primary component. 

Interestingly enough, \citet{Gizis02} note that in close proximity to the target source, only 5.9 arcseconds south and 3.3 arcseconds west is another object, 2MASS J15104761-2818234, with similar spectral type of M8-M9. The close proximity could indicate for the 2 objects to be part of the same system, and more recent Gaia DR2 results show similar trigonometric parallax values for both of the late M-dwarfs, along with proper motions, suggesting that they are indeed likely to be part of the same system. The projected separation of the 2 objects and the parallax suggest that they have a physical separation of about 250 AU. Furthermore, Triaud et al.\footnote{ESO Proposal ID 299.C-5046(A) --  \url{http://archive.eso.org/wdb/wdb/eso/abstract/query?ID=9950460&progid=299.C-5046\%28A\%29}} find through photometric monitoring of the system that 1 of the components is an eclipsing binary, making this a very interesting system from a hierarchical point of view, with perhaps 4 substellar components. This scenario could partly explain the discrepancy in temperature for the primary and binary components detected in our observations. It is possible that 2M1510A itself is a close-in multiple system which combined brightness gives the appearance of an overluminous object with respect to its temperature, similar to other unresolved substellar multiples such as 2M1610-1913 \citep{Lachapelle+15}.

\subsubsection{2MASS J15474719-2423493}
Previously identified as a M9$\beta$ spectral type low-gravity object with an age estimate ranging $30 - 50$ Myr, corresponding to a $12.9 \pm 0.3\,M_{\rm Jup}$ planetary-mass object \citep{Gagne+15}. Updated space velocities and parameters from Gaia DR2 with proper motions $\mu_{\rm RA} = -141.12 \pm 0.72$ mas/yr, $\mu_{\rm DEC} = -132.71 \pm 0.46$ mas/yr and parallax $\pi = 29.20 \pm 0.40$ mas  combined with the updated BANYAN $\Sigma$-tool suggests that the system belongs to the field rather than a YMG. \citet{Gagne+15} evaluate the system as a Young Field Object based on several signs of youth, which includes; lower than normal equivalent width of atomic species in the optical spectrum as well as in the NIR spectrum, triangular-shaped $H$-band continuum, and redder than normal colours for the given spectral type. The Convergence tool suggests a probability of 31.3$\%$ for 2M1547 to be in the Columba YMG, which has an estimate age of $\approx 35$ Myr and thus consistent with the other age estimates for the system. 

We identify a faint companion to the main source with a projected separation of $173.66 \pm 0.50$ mas, seen in the middle plot of Figure~\ref{fig:binaries}. Using the conversion factor of 1.16 for converting the physical separation to a statistical estimate of the semi-major axis we obtain $a \approx 7.0$ AU, translating into an orbital period of $P \approx 100$ years. However, the observations are of poor quality and the signal for either components identified are too faint to be seen in individual wavelength-slices, and only visible when collapsing the data cube. The dithering pattern for the observations of 2M1547 is very small, sometimes less than a pixel, and thus we cannot obtain good sky subtraction for the observations without self-subtraction of the source. To work around the issue we perform the reduction scheme with the SINFONI pipeline without any sky subtraction, and post process each individual reduced frame by subtracting sky-frames from observations taken just before and after the observations of 2M1547. The improvement is small, but allows us to identify the companion candidate with higher signal-to-noise ratio than using no sky subtraction at all.

We perform our astrometric analysis on the reduced and post-processed collapsed frames, obtaining projected separations, positional angles and relative brightness between the components. We estimate the secondary source in the field to be about a factor of $\approx 0.55$ as bright as the primary source in the $H'$-band. For the $K'$-band we obtain similar brightness for both components, which is likely due to the poor quality of the observations and weak signal for both sources in the $K'$-band. As with the other binaries, we estimate the individual component masses using the BT-Settl-CIFIST evolutionary models of \citet{Baraffe+15}. For 2M1547 we obtain a primary mass of $M_{\rm primary} = 19.51^{+4.48}_{-2.30}\,M_{\rm Jup}$ and secondary mass of $M_{\rm secondary} = 15.51^{+3.85}_{-2.05}\,M_{\rm Jup}$ for the adopted young field age of $30-50$ Myr, corresponding to a mass-ratio of $q = 0.7 - 0.8$.

We are unable to obtain high signal for the spectra retrieved from the sources in the final image of the reduced data cube for this particular target. Nevertheless, from the data we are able to see a higher signal in the $H'$-band with respect to $K'$-band, and features resembling that of brown dwarf spectra for both the brighter source and fainter source in the image. Due to the faint nature of the sources and the poor quality of the spectra obtained for 2M1547, we cannot make a clear distinction for the spectral types of the sources and thus cannot rule out the possibility that they are background sources. Nevertheless, given the small separation between the 2 components in the image frame, it is very likely that they are indeed part of the same system, and we calculate a chance projection of a background source of $5.7 \cdot 10^{-5}$. We encourage follow-up observations of the target that may clarify whether 2M1547 is a bona-fide young planetary-mass binary or not.

Incidentally, after applying the reduction from the development-level pipeline, correcting for atmospheric dispersion, we identify a third point-like source in the datacube that was not clearly visible in previous reduction attempts. The source is seen in the upper left of the upper middle image of Figure~\ref{fig:binaries}, separated from the primary component by $\sim 220$ mas and with a positional angle of $\sim 73^{\circ}$. We estimate the relative brightness of this potentially tertiary component as $\approx 0.40$ of the brightness of the primary. Nevertheless, due to the ambiguous nature of the source and the issues with the data set for 2M1547, we do not include it in our final analysis and instead consider only the binary candidate.

\subsubsection{2MASS J22025794-5605087}
\begin{figure*}[hbtp]
  \centering
  \includegraphics[width=\columnwidth]{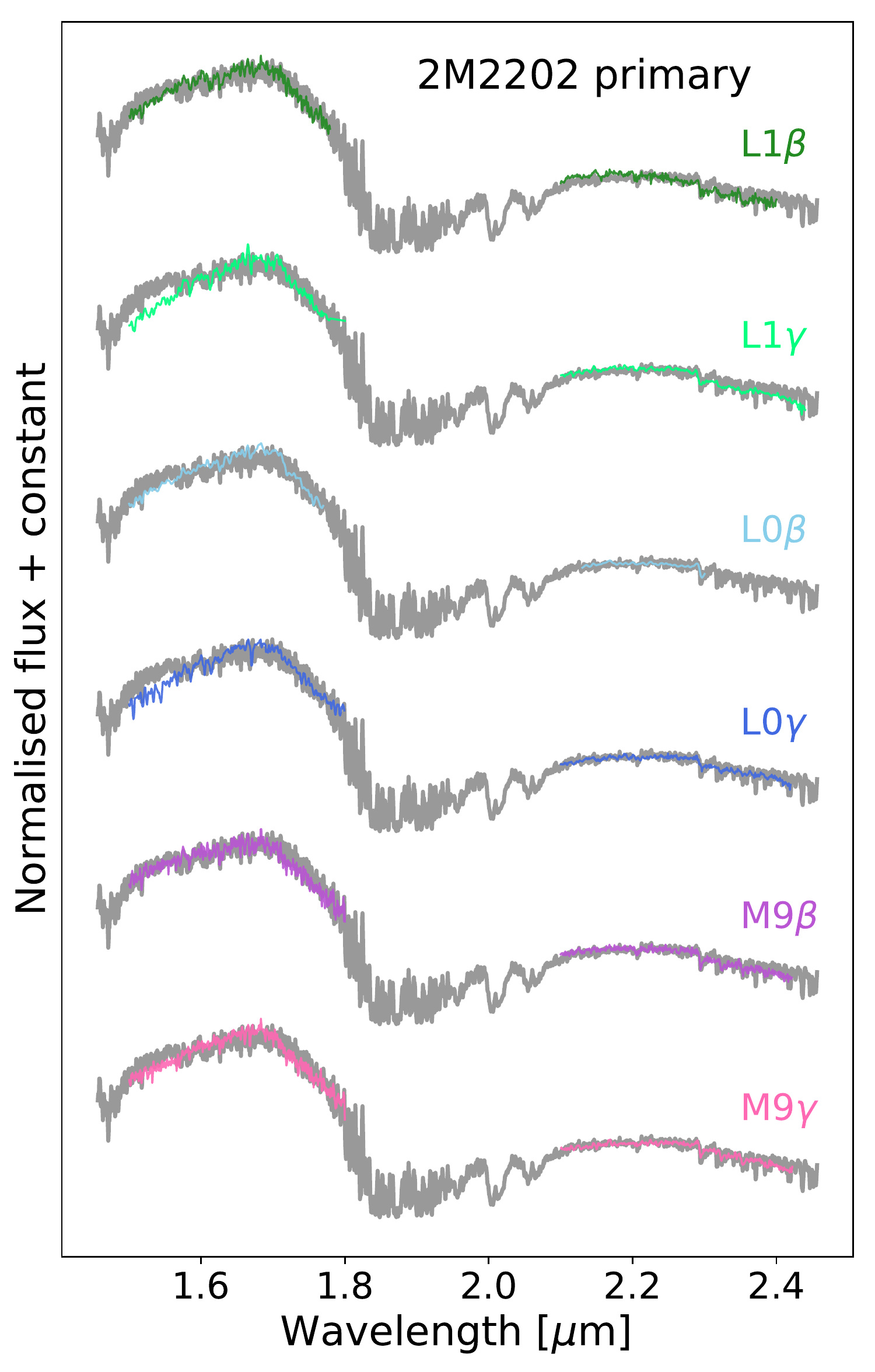}
  \includegraphics[width=\columnwidth]{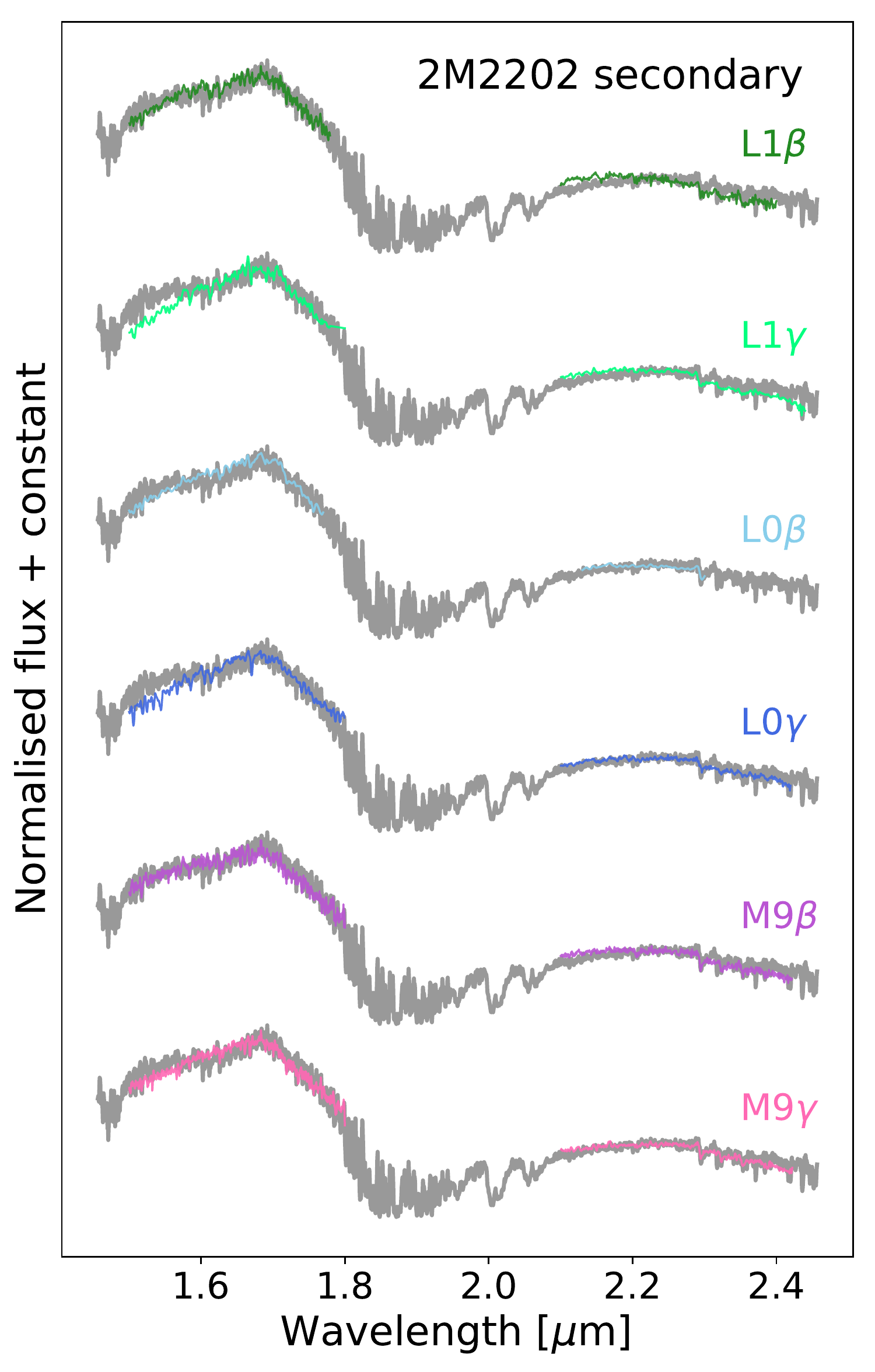}
 \includegraphics[width=\columnwidth]{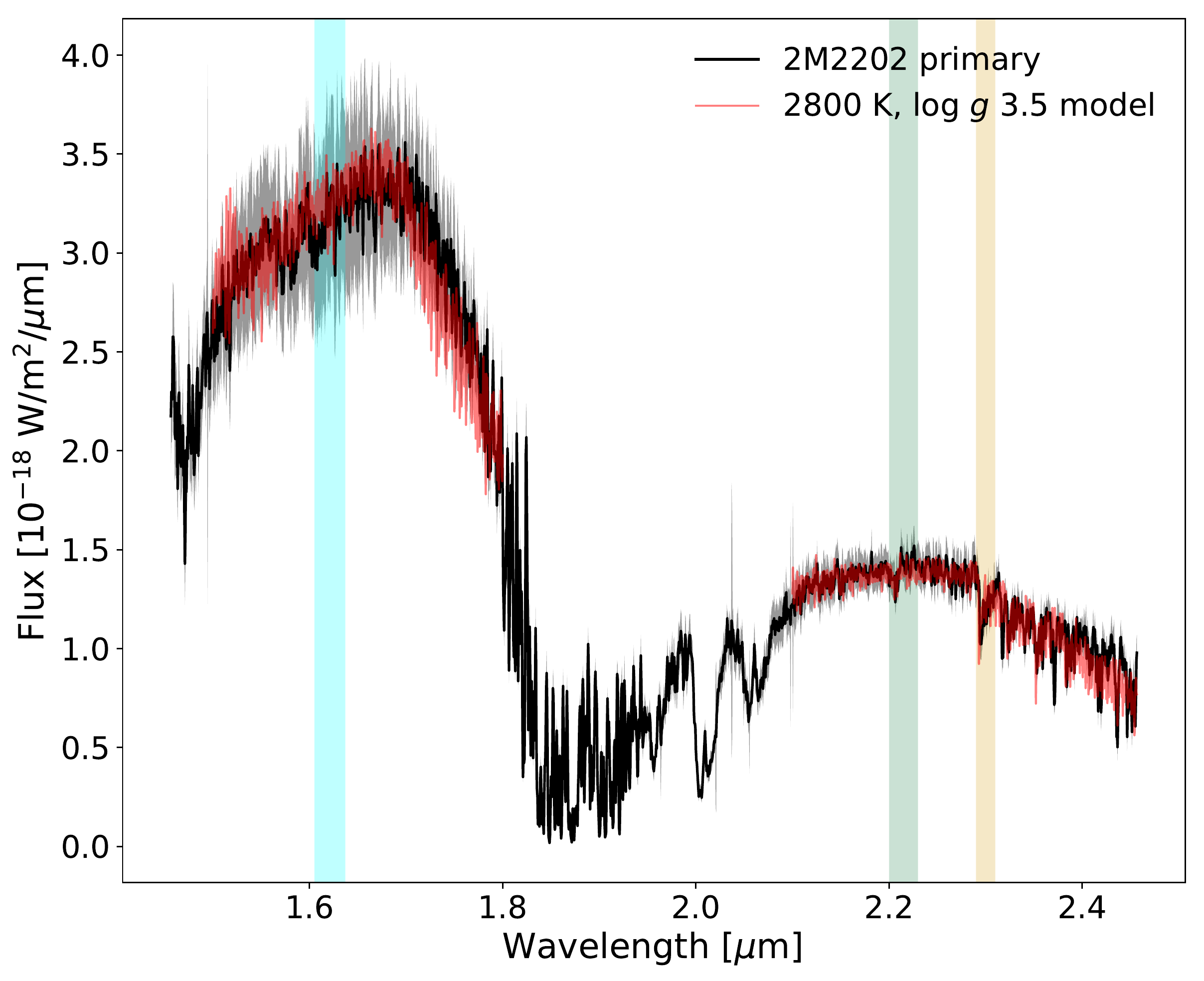}
 \includegraphics[width=\columnwidth]{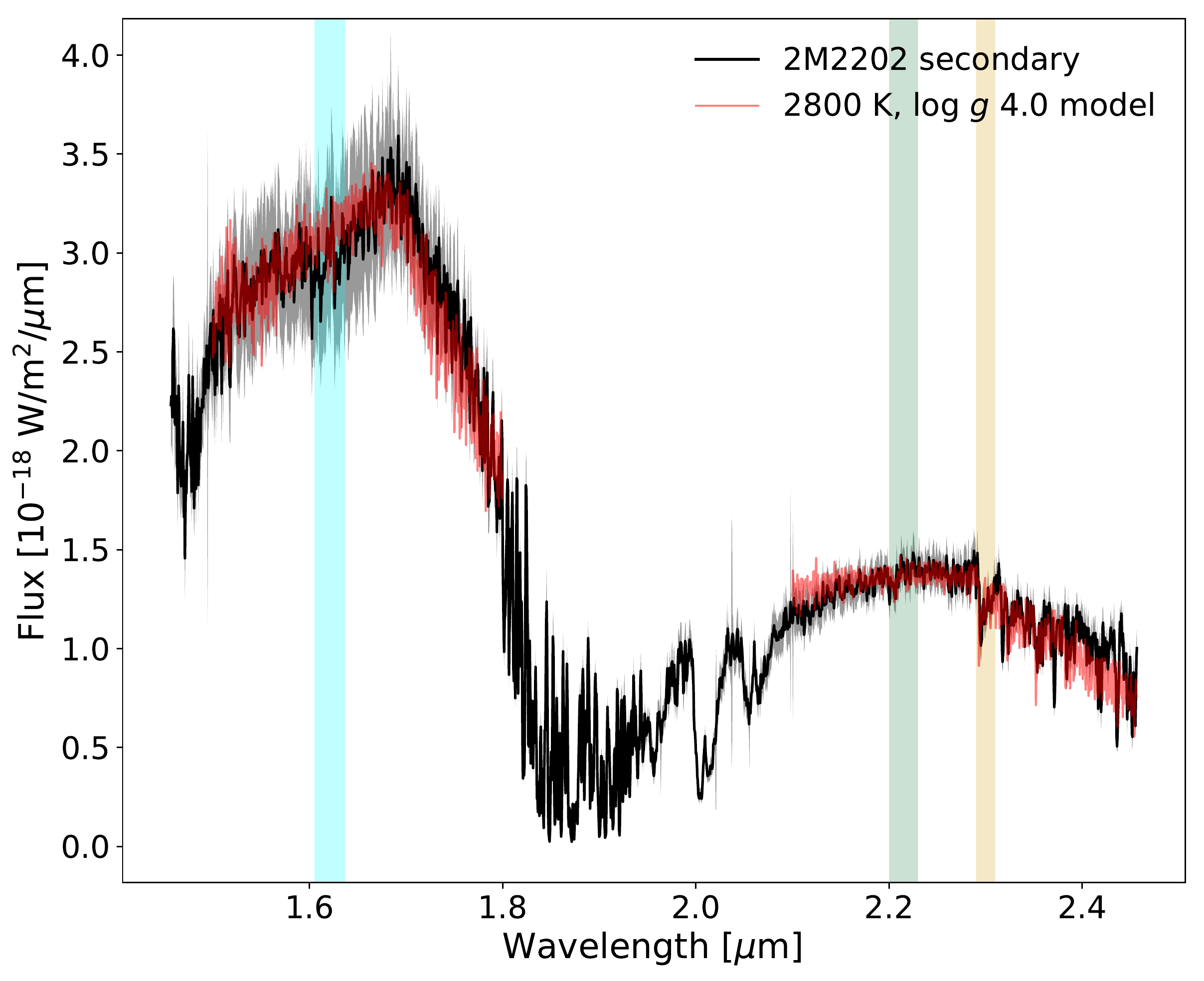} 
  \caption[]{\label{fig:2M2202} %
Top row depicting empirical templates fitted to 2M2022. We find a best-fit model to be of M9$\gamma/\beta \pm 1$ for both components, suggesting for a pair of young brown dwarfs around $\sim 100$ Myr old. Due to the small relative separation between the 2 components we expect there to be quite some contamination from each of them influencing the other. Nevertheless, we find that they are of almost equal brightness and expect them to share spectral type. The spectral type is consistent with the age of the AB Dor Moving Group at around 120-200 Myr, suggesting that the binary is a pair of $\approx 50\,M_{\rm Jup}$ brown dwarfs. The lower plots show the best-fit theoretical evolutionary models from BT-Settl CIFIST to each individual component, where we find the $T_{\rm eff} = 2800$ K and $\log g = 3.5 - 4.0$ models to be the best fit to both components. The light-blue, green and yellow shaded areas indicate the spectral features from FeH, Na I and CO respectively.
   }
\end{figure*}

The system has previously been identified as an M9 spectral type, with space velocities placing it as a high probability member of the Tucana-Horologium (THA) moving group \citep{Gagne+15}. More recent results using updated Gaia DR2 parameters with proper motions $\mu_{\rm RA} = 52.64 \pm 0.66$ mas/yr, $\mu_{\rm DEC} = -77.13 \pm 0.73$ mas/yr and parallax $\pi = 14.29 \pm 0.48$ mas with the BANYAN $\Sigma$ tool suggest 2M2202 to be in the somewhat older AB Dor (ABD) moving group, which is consistent with the Convergence tool as well. Our spectral analysis of the individual components in the system, seen in Figure~\ref{fig:2M2202}, implies both to be consistent with M$9\gamma$ and M$9\beta$ spectral types, hinting that they are more than 100 Myrs old, which agrees with the system belonging to the ABD moving group.

Visual inspection of the collapsed wavelength cube in the rightmost plot in Figure~\ref{fig:binaries} clearly shows 2 resolved individual components of almost equal brightness, with the southern component being slightly fainter, which we will refer to as the binary component or 2M2202B. We calculate a chance projection of the fainter object being a background source with probability of $2.4 \cdot 10^{-6}$. Our astrometric solution, shown in Table~\ref{tab:results}, reveals the binary to have a projected separation of $60.36 \pm 2.67$ mas and positional angle of $59.51 \pm 1.58$ deg. We estimate a physical separation for the components by dividing the projected separation by the Gaia DR2 trigonometric parallax, and convert the separation to a statistical argument for the semi-major axis as with the other binaries. The predicted semi-major axis for the 2M2202 binary becomes $a \approx 4.8$ AU, corresponding to an orbital period of $P \approx 34$ years.

From the resolved photometry we fit the luminosity of each component to isochrones from the BT-Settle CIFIST models which yield primary and secondary masses of $M_{\rm primary} = 57.37^{+11.30}_{-7.90}\,M_{\rm Jup}$ and $M_{\rm secondary} = 55.60^{+10.89}_{-7.43}\,M_{\rm Jup}$ respectively for the ABD MG age range of $120 - 200$ Myr. Should the system in fact belong to the THA moving group instead, which has an age range between $20 - 40$ Myr \citep{Kraus+14}, the estimated masses change to the mass-range of $\approx 15-29\,M_{\rm Jup}$. Regardless of adopted age, the mass ratio for the system is close to unity. The previous mass-estimate by \citet{Gagne+14} of $22.1^{+4.7}_{-3.9}\,M_{\rm Jup}$ for the unresolved system employs similar evolutionary models as here but are likely to be underestimated as it utilizes a statistical distance of $d_{\rm statistical} = 48.6$ pc instead of the more precise Gaia DR2 trigonometric parallax distance of $d_{\rm Gaia} = 70.0$ pc. The difference in distance corresponds to a shift of about 0.8 in absolute magnitudes, which in combination of the earlier adopted younger age of THA explains the previous lower mass-estimate.

The spectral analysis for the individual components for 2M2202 indicate both to be of M9$\beta$ spectral type, which is expected due to their similar brightness and assumed age. However, the theoretical modelling is more degenerate, and visualised in the lower plots of Figure~\ref{fig:2M2202}. We find best-fit models with effective temperatures of  $2800 \pm 100$ Kfor the primary and secondary both, with surface gravities of $\log g = 4.0 \pm 0.5$. The theoretical spectra predicts higher effective temperatures for the individual components compared to the isochrones from the same models, showing a discrepancy of about 400 K. Although the $T_{\rm eff} = 2400$ K models fit well with observed slope in the $K'$-band, in the $H'$-band the fit is worse compared to the higher temperature models.

\subsection{Singles or unresolved sources} \label{sec:singles}
The following sections contain more detailed information regarding the objects identified as single or unresolved substellar objects with no close-in companions. The same spectral analysis is performed for these objects as with the binaries, and we compile the empirical model comparison for all single objects in Figure~\ref{fig:singlespecs}.

\subsubsection{2MASS J20334473-5635338}
\begin{figure*}[hbtp]
  \centering
  \includegraphics[width=180mm]{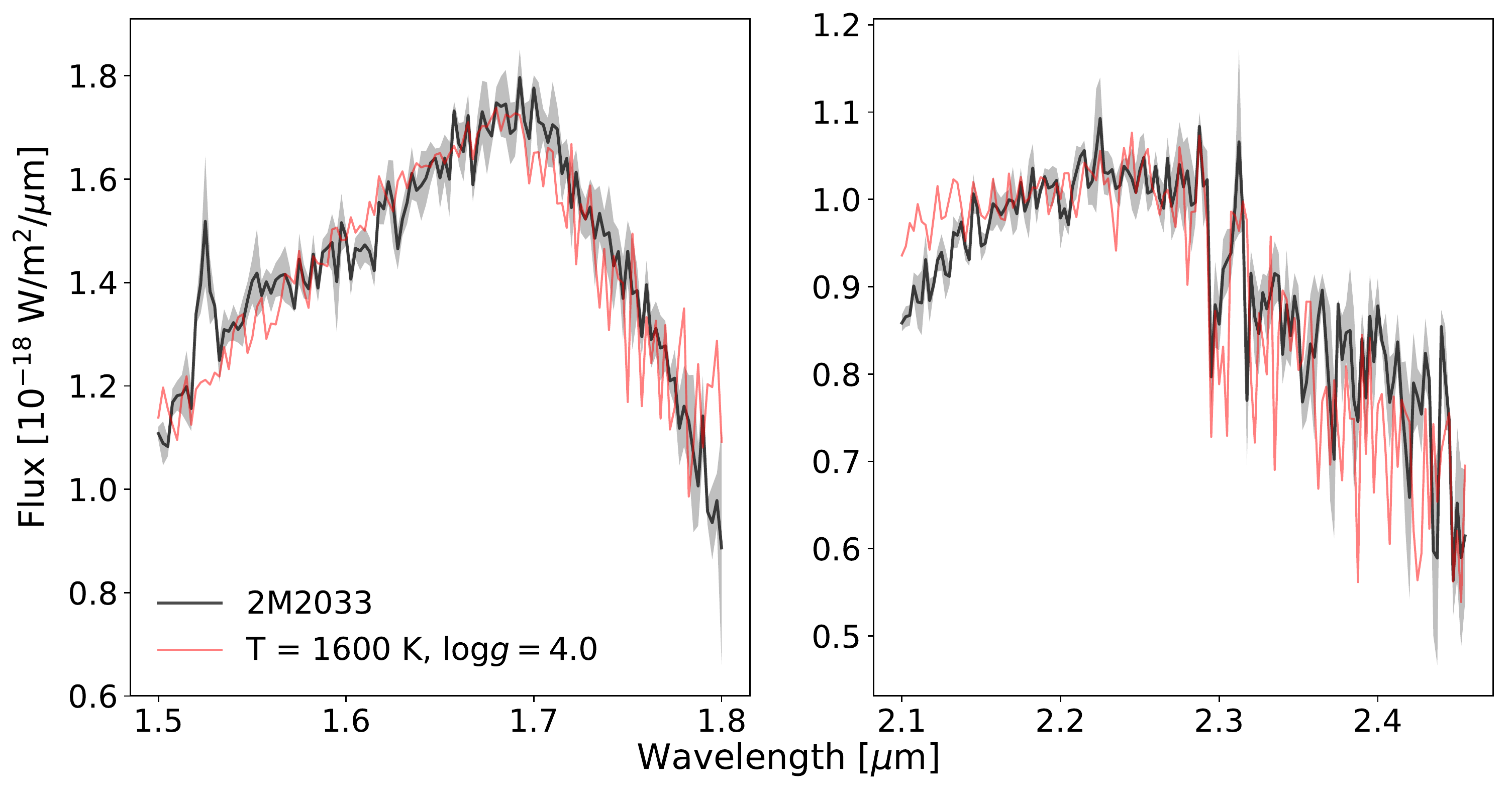}
  \includegraphics[width=160mm]{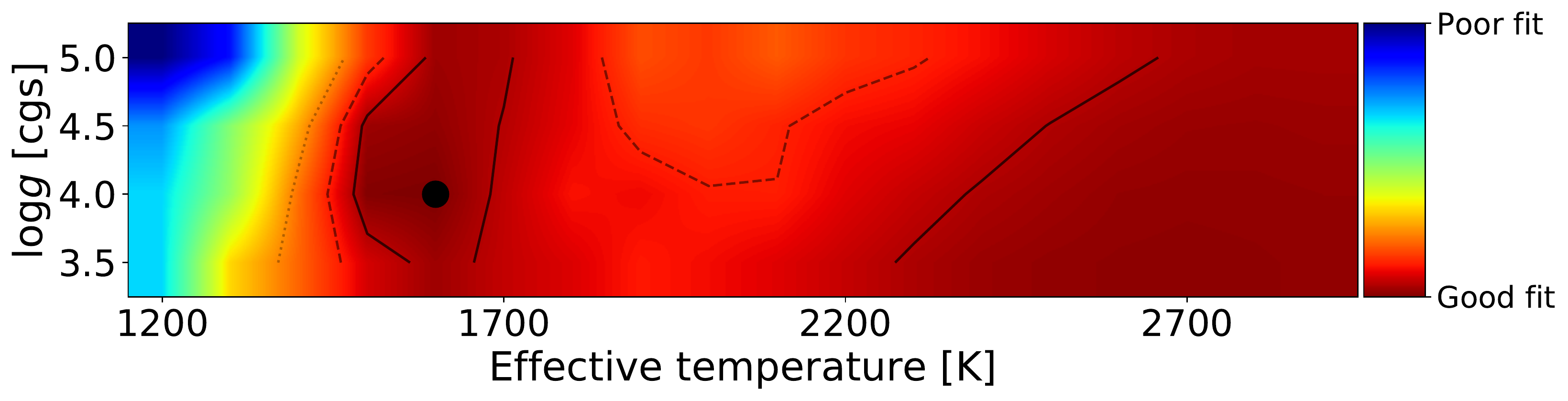}  
  \caption[]{\label{fig:2M2033} %
  Comparison for spectra of 2M2033 and theoretical models from the BT-Settl CIFIST library. The upper panels depict the observed flux in black where we have binned the observed flux into wavelength-slices of 5 for a better signal-to-noise ratio, obtaining a wavelength resolution of $\Delta \lambda = 25\,\AA$, and using the standard deviation from the mean of the bins as the uncertainty, shown by the grey-shaded areas. The red lines depict the best-fit theoretical model, here with $T_{\rm eff} = 1600_{-100}$ K and $\log g = 4.0_{-0.5}^{+1.0}$. The lower colour-map show the relative goodness-of-fit values for the different models tested, where the blue represent poor fits and the red colour represent good fits, and the best fit with uncertainty included is illustrated by the black marker and ellipse. 
   }
\end{figure*}

We classify 2M2033 as an L0$\gamma$ dwarf and a likely member of the THA moving group. Our spectral analysis confirms the previous results that it is most likely a young L0 dwarf, and we obtain a best-fit theoretical model with effective temperature of $T_{\rm eff} = 1600_{-100}$ K and surface gravity $\log g = 4.0_{-0.5}^{+1.0}$, as shown in Figure~\ref{fig:2M2033}. The object has a statistical distance of about 55 pc \citep{Gagne+15}, but is lacking trigonometric parallax from Gaia DR2. Previous mass-estimate by \citet{Gagne+15} indicates $13.5\,M_{\rm Jup}$ for the object, placing it close to the classical Deuterium burning limit and making it a potential planetary-mass substellar object.

\subsubsection{2MASS J21171431-2940034}
\begin{figure*}[hbtp]
  \centering
  \includegraphics[width=180mm]{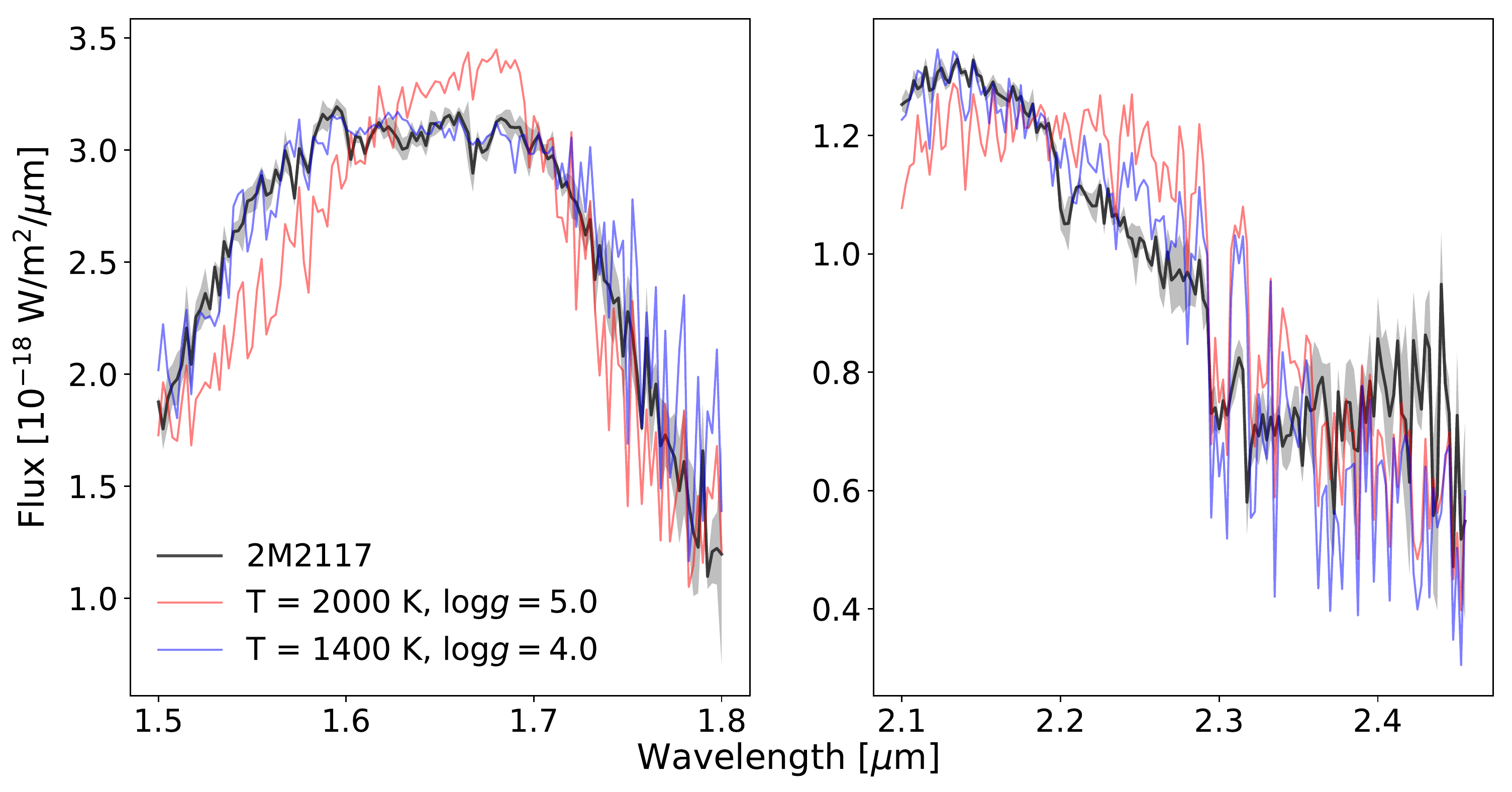}
  \includegraphics[width=160mm]{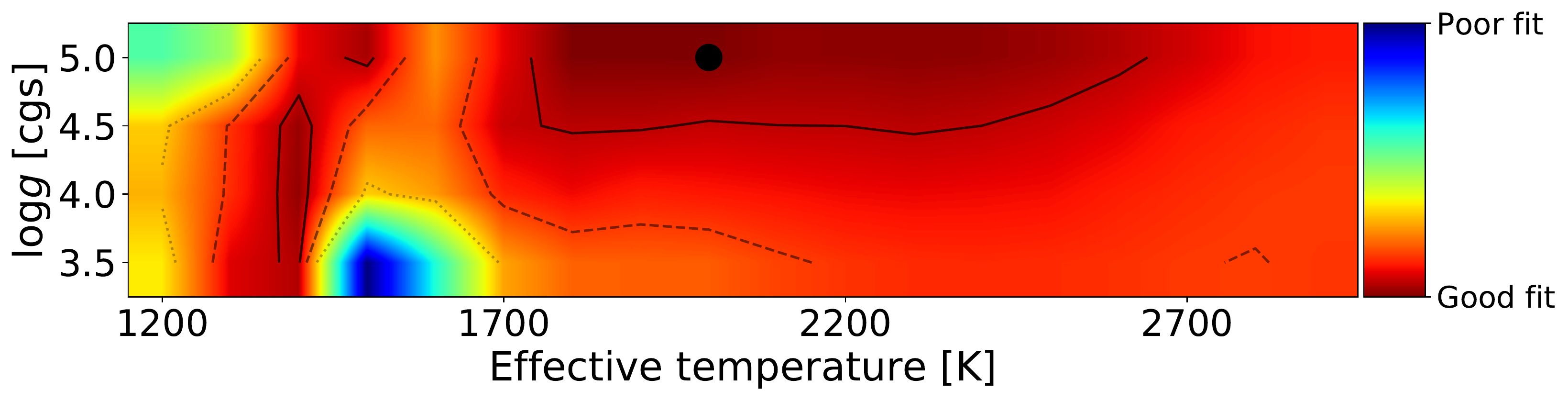}  
  \caption[]{\label{fig:2M2117} %
  Comparison of observed spectra for 2M2117 and theoretical models from the BT-Settle CIFIST library, same as in Figure~\ref{fig:2M2033}, but with the best-fit model being $T_{\rm eff} = 2000^{+600}_{200} $ K with $\log g  = 4.5_{-0.5}$. The theoretical models show strong signs of degeneracy when we match them to the target, and we obtain better fits for unlikely models with much higher temperature than what is expected for the given spectral type of T0. The blue line in the upper plots depict a lower temperature model with $T_{\rm eff} = 1400$ K and $\log g = 4.0$ with similar goodness-of-fit statistic as the higher temperature best-fit model.
   }
\end{figure*}

Single T0-type object with a distance of about $16.9 \pm 1.7$ pc \citep{Best+15} and likely member of the $\beta$ pic moving group. Previous mass-estimate suggest for a $6.5\,M_{\rm Jup}$ planetary mass object. Our spectral analysis finds a best-fit for an effective temperature of $T_{\rm eff} = 2000$ with a surface gravity of $\log g = 5.0$. However, as indicated by the theoretical spectral goodness-of-fit map shown in the bottom panel of Figure~\ref{fig:2M2117}, the theoretical models are highly degenerate for this object and there are several possible matches. Comparing isochrones from the same evolutionary models as employed in the spectral fit, we estimate from the brightness of 2M2117 that the effective temperature should be closer to $\sim 1000$ K, which is in better agreement with the fitted spectral type of T0. We see from the grid-map in the lower plot in Figure~\ref{fig:2M2117} that lower temperature models than our best fit are within the $1-\sigma$ level from the goodness-of-fit statistics. Hence, we include the best-fit model from the lower temperature plateau in the figure, corresponding to an effective temperature of $T_{\rm eff} = 1400$ K with a surface gravity of $\log g = 4.0 \pm 0.5$.

\subsubsection{2MASS J22353560-5906306}
\begin{figure*}[hbtp]
  \centering
   \includegraphics[width=180mm]{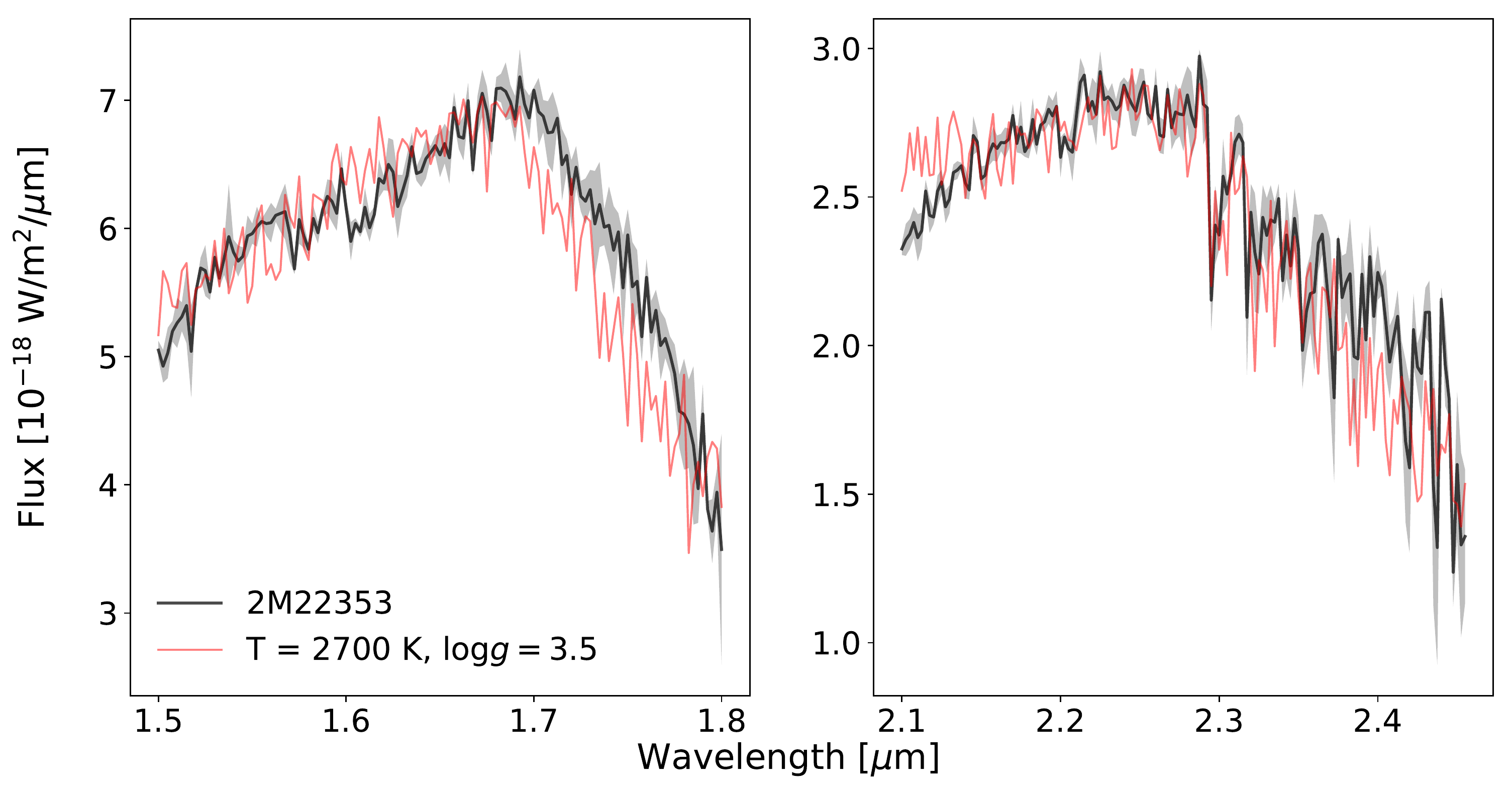}
    \includegraphics[width=160mm]{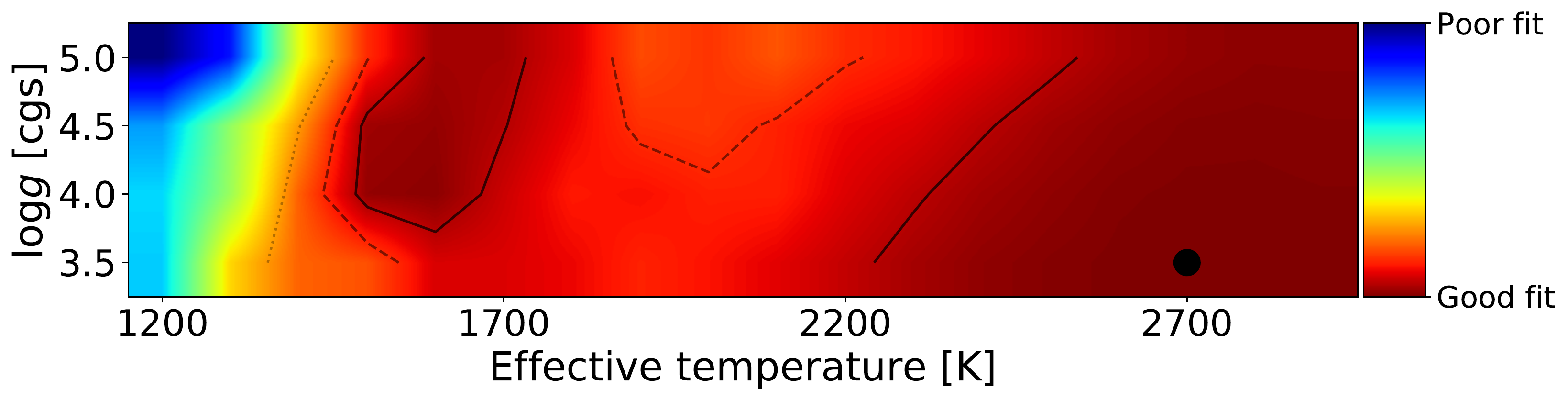}
  \caption[]{\label{fig:2M22353} %
  Comparison of observed spectra for 2M22353 and theoretical models from the BT-Settle CIFIST library, same as in Figure~\ref{fig:2M2033}, but with the best-fit model being $T_{\rm eff} = 2700 \pm 400$ K with $\log g  \leq 3.5$. 
   }
\end{figure*}

Likely member of the THA moving group and previously classed as a young M9$\beta$ spectral type dwarf with an estimated mass of $\approx 20.6\, M_{\rm Jup}$ \citep{Gagne+15}. Gaia DR2 shows proper motions of $\mu_{\rm RA} = 59.37 \pm 0.39$ mas/yr and $\mu_{\rm DEC} = -83.76 \pm 0.43$ mas/yr, with a trigonometric parallax pointing to a distance of $\approx 46$ pc for the object \citep{Lindegren+18}. Our spectral analysis finds a best-fit spectral type of M9$\gamma$ for the source and a theoretical model close to an effective temperature of $T_{\rm eff} = 2700 \pm 400$ K and surface gravity of $\log g \geq 3.5 $, shown in Figure~\ref{fig:2M22353}. We note that there is a huge discrepancy in effective temperature of $\approx 1000$ K and magnitude of $\approx 1.5$ mag between the sources 2M2033 and 2M22353, despite them having similar spectral types of M9 and L0 respectively. This oddity is enhanced by the fact that both of these objects are likely members of the THA young moving group, albeit it is not unheard of for brown dwarfs of similar spectral types to have dissimilar temperatures and brightness \citep[e.g. 2M1610-1913][ or 2M1510 here]{Lachapelle+15}.

\subsubsection{2MASS J23225299-6151275}
\begin{figure*}[hbtp]
  \centering
  \includegraphics[width=180mm]{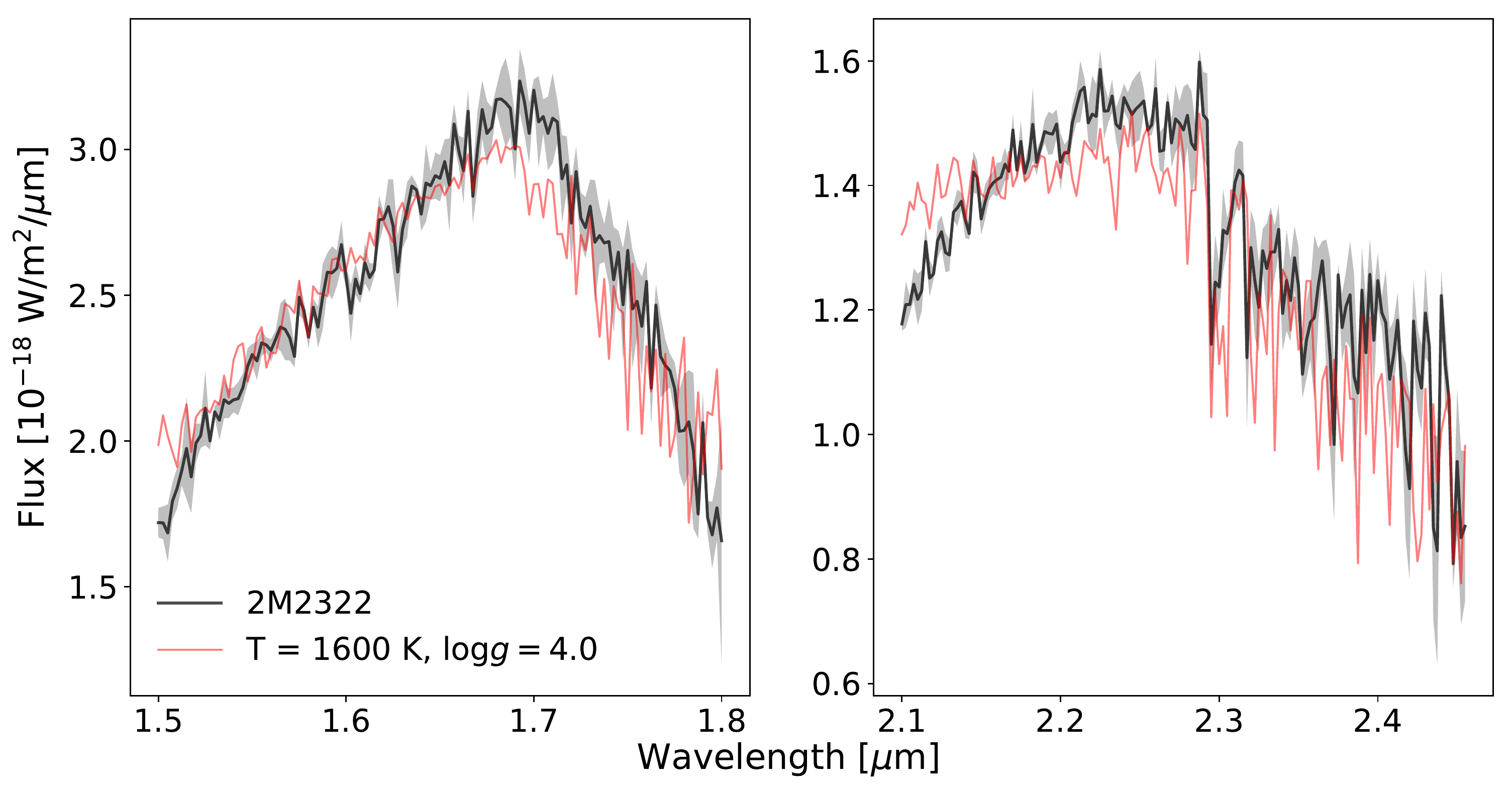}
  \includegraphics[width=160mm]{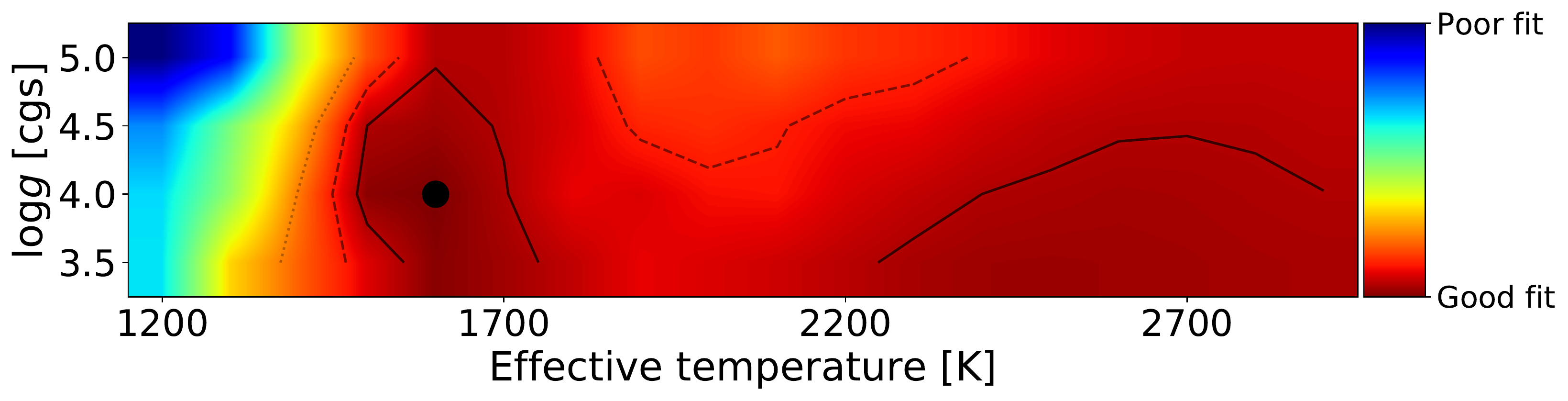}  
  \caption[]{\label{fig:2M2322} %
  Comparison of observed spectra for 2M2322 and theoretical models from the BT-Settle CIFIST library, same as in Figure~\ref{fig:2M2033}, but with the best-fit model being $T_{\rm eff} = 1600 \pm 100$ K with $\log g  = 4.0 \pm 0.5$. 
   }
\end{figure*}

The target is potentially part of a wider binary system together with the low-mass M5 type dwarf 2MASS J23225240-6151114 \citep[e.g.][]{Gagne+15}, sharing similar trigonometric parallax and proper motions according to the Gaia DR2 release. The target was previously determined to be a L1$\gamma$ BD by \citet{Gagne+15}, but later estimated by \citet{Koen+17} to be of slightly later type as an L2.5 BD. We find similar results and obtain good fits for L1$\gamma \pm 2$ spectral templates. We find theoretical models to be best fitted with an effective temperature of $T_{\rm eff} = 1600 \pm 100$ K and with a surface gravity of $\log g = 4.0 \pm 0.5$, shown in Figure~\ref{fig:2M2322}. Using the Gaia DR2 proper motions of $\mu_{\rm RA} = 80.09 \pm 1.45$ mas/yr and $\mu_{\rm DEC} = -81.97 \pm 1.62$ mas/yr with the BANYAN $\Sigma$ tool associates 2M2322 to the THA moving group, and the trigonometric parallax of $\pi = 23.24$ mas places it at a distance of $\approx 43$ pc, yielding an isochronal mass of $\approx 13\,M_{\rm Jup}$.

\section{Discussion \& Summary} \label{sec:disc}
The combination of the small working angle of the observations and the resolved spectra lets us place strong constraints for 2M1510 and 2M2202 being binary BDs. They both show strong signs of youth and share space velocities with the Argus YMG and ABD MG respectively, which in combination of assuming that all members of the YMGs are coeval can constrain their ages even further to $30 - 50$ Myr for 2M1510 and $120 - 200$ Myr for 2M2202. We also find a tentative binary BD in the young field, 2M1547, predicted to have an age of $30 - 50$ Myr according to isochronal fits of the near-IR photometry. The main reason for 2M1547 only being a binary candidate comes from the poor quality of the observations of the target, where the source appeared to be much fainter than expected based on 2MASS photometry. We are not certain why the observations of 2M1547 were of less quality than expected, but it is possible that thin local clouds got in the way at the crucial moment of the exposures. Nevertheless, we verify that the coordinates for the observations are indeed correct and that no other source of similar faint nature should be present in the vicinity. The observations also lacked proper sky-subtraction, to which we resort to subtract post-processed sky frames from observations just before and after 2M1547 was observed. However, we calculate a likelihood for chance projection that the detected binary candidate components are not being physically bound to be on the order of $10^{-5}$ and therefore almost negligible. We summarise the observations and YMG membership status in Table~\ref{tab:obs} for all observed targets.

Using standard stars observed on the same nights as our targets, we employ a PSF-matching algorithm to model the relative brightness and positions of the resolved components of the binaries. From the relative positions of the detected companions we deduce the projected separation of the binaries to be $\approx 100$ mas for 2M1510 and $\approx 60$ mas for 2M2202, corresponding to physical separations of $\approx 4 - 5$ AU for the adopted distances. The binary candidate to 2M1547 has a larger separation of $\approx 170$ mas, corresponding to a physical separation of $\approx 6.0$ AU. These small separations can be related to the expected period of the binaries, which translates to a few decades for 2M1510 and 2M2202, and about a 100 years for 2M1547. We summarise the astrometry for the binaries in Table~\ref{tab:results}.

From the relative brightnesses of the binary components and comparing their ages to isochrones from the BT-Settle CIFIST \citep{Baraffe+15} evolutionary models we predict individual masses. We obtain that 2M1510 is likely composed of a $M_{\rm 2M1510A} = 41.50^{7.04}_{-7.37}\,M_{\rm Jup}$ primary and a secondary BD with $M_{\rm 2M1510B} =17.68^{+4.20}_{-2.10}\,M_{\rm Jup}$, or perhaps even lower in mass, as the PSF is somewhat smeared out and we are likely to obtain a significant brightness contamination contribution from the primary. We obtain that 2M2202 is an almost equal-mass binary, with the primary component mass as $M_{\rm 2M2202A} = 57.37^{+11.30}_{-7.90}\,M_{\rm Jup}$ and the secondary with $M_{\rm 2M2202B} = 55.60^{+10.89}_{-7.43}\,M_{\rm Jup}$. We find the total mass of 2M2202 to be higher than previous results from \citet{Gagne+15} due to the updated distance measurements and space velocities from Gaia, thus constraining its YMG membership and adopted age. For the binary candidate 2M1547 we predict masses of $M_{\rm 2M1547A} = 19.51^{+4.48}_{-2.30}\,M_{\rm Jup}$ and $M_{\rm 2M1547B} = 15.51^{+3.85}_{-2.05}\,M_{\rm Jup}$, and both of the components are close to the classical Deuterium burning limit, making it an interesting binary system from an evolutionary perspective. We compile the photometric and mass results in Table~\ref{tab:results}, where we predict the masses using isochrones from the evolutionary models and the ages stated in table. We note that using the lower ages of the YMGs would lower the masses further, but in some cases the models break down at these very low masses and ages. There may also be an issue for dusty and cloudless substellar evolutionary models to correctly predict cooling rates as \citet{Dupuy+14} point out, suggesting that model-derived masses may be overestimated. 

Here, we keep our calculations simplified by applying the same models for the isochrones as for the theoretical spectral analysis but note that different evolutionary models may yield divergent results. We also notice some discrepancy in the comparison between isochrones and spectral fitting, where the best-fit spectra to our objects produce higher effective temperatures than the isochrones, but lower surface gravity. We can attribute these discrepancies partly to assuming solar metallicities for the models, which is not necessarily the case for our young objects; \citet{VA+09} find for instance that the ABD MG has somewhat supersolar metallicty of $[{\rm Fe}/{\rm H}]_{\rm ABD} = 0.04 \pm 0.05$; the Argus group to have $[{\rm Fe}/{\rm H}]_{\rm ARG} = 0.02 \pm 0.06$; and THA to have $[{\rm Fe}/{\rm H}]_{\rm THA} = -0.03 \pm 0.05$. Furthermore, we note a strong degeneracy with the theoretical models when attempting to match the models to the observed spectra. By alternating effective temperatures and surface gravities we achieve very similar values in our goodness-of-fit calculations and shapes in the spectra. For the 2M1510 binary we can contribute some of this degeneracy to the only partly resolved spectra, where the brightness of the primary clearly contaminates the secondary component and its spectrum. We also cannot rule out the possibility of unresolved components of different spectral types that causes models of very different values of $T_{\rm eff}$ and $\log g$ to produce good fits.

We construct empirical templates for different spectral types using a selection of known objects, as well as using already existing templates to try to constrain the spectral types of our targets and discovered binaries. In most cases we robustly verify previous indications, and for the partly resolved 2M1510 binary candidate the spectrum of the companion is too contaminated to yield constrained results. We do note however, that the empirical models for the binary companion agree well with a later spectral type than for the primary.

Assuming that the targets in our observations are true members of their respective YMG, and that all members are coeval, we detect 2 out of 6 YMG substellar objects to be binaries, with the addition of 1 strong binary candidate from the young field population of substellar BDs. The sample provided here is too small to set stringent constraints on the multiplicity for young low-mass substellar objects. We do however provide 3 new binary systems in this scarcely probed regime of mass and young objects, along with 4 likely single systems that may also be used for future multiplicity studies. The very narrow field of view of the observations taken here of 0.8''$\times$0.8'' is only viable for searching for binaries on small separations, which these type of objects are expected to constitute. Thus, from our sample of 7 observed targets we find 3 of them to be binary candidates, leading to a multiplicity frequency of $\approx 43 \pm 19\,\%$ for BDs with projected separations effectively smaller than $\approx 400$ mas. Such number has been speculated for the overall multiplicity rate of BD-BD pairs based on color-magnitude diagrams and radial velocity variation statistics \citep{Pinfield+03, MJ05}. Although we are dealing with a very small sample, we cannot rule out that this could have some indication towards young systems in this mass-range to have a higher multiplicity frequency than their older field counterpart. This could also be related to the narrow separations which would have been difficult to detect in other surveys.

We identify 2M1510 as a close-in binary pair, which is of particular interest from a dynamical and hierarchical point of view, already being in a well-separated binary system with at least one of the components having an eclipsing binary and another close-in companion. Such circumbinary systems would be of interest for studying formation channels and dynamical evolution of young BD-BD systems, which can be compared to previous multiple studies of more massive close-in binaries with circumbinary companions \citep[e.g.][]{Ruben+18}. The fact that our binary candidates are all on small separations supports the formation scenarios of BDs as a substellar extension of the low-mass stellar initial mass-function \citep[e.g.][]{GW07, Luhman12}. In order to test a wider range of formation scenarios, multiplicity studies of even younger systems than observed here, for example in star-forming regions such as Taurus, Chamaeleon I or Scorpius-Centaurus would be required.

The small separations for the 3 candidate binary systems detected here are on the order of a few AU, which combined with their predicted masses from substellar evolutionary models yield Keplerian orbital periods of just a few decades for the most close-in binaries 2M1510 and 2M2202. Our spectral analysis suggests for an unambiguous spectral class of M$9\beta$ for both components to 2M2022, implying that the system is indeed a bona-fide substellar binary. Based on the previous discussion in Section~\ref{sec:2M1510}, we deem 2M1510 to be a strong BD binary candidate on the basis that the shape of the spectra are slightly different and that a point source is identified after subtracting a model PSF from the bright main component. Nevertheless,  we advocate for second-epoch follow-up observations of 2M1510 and 2M1547 in order to verify common proper motions and their binarity status. Given small separations and periods of just a few decades, Keplerian orbits can be fitted within just a few years of monitoring of these 3 systems, which typically can be done after just monitoring $\approx 1/3$ rd of the orbit, thus yielding dynamical masses within a relatively short time. Since the ages are already constrained from the YMGs, these systems will be excellent benchmarks for calibrating theoretical models in this very low mass-regime.


\begin{acknowledgements} 
 M.J. gratefully acknowledges funding from the Knut and Alice Wallenberg foundation.
This research has benefitted from the SpeX Prism Library (and/or SpeX Prism Library Analysis Toolkit), maintained by Adam Burgasser at http://www.browndwarfs.org/spexprism". This research has benefitted from the Montreal Brown Dwarf and Exoplanet Spectral Library, maintained by Jonathan Gagn\'{e}. This work has made use of data from the European Space Agency (ESA) mission {\it Gaia} (\url{https://www.cosmos.esa.int/gaia}), processed by the {\it Gaia} Data Processing and Analysis Consortium (DPAC, \url{https://www.cosmos.esa.int/web/gaia/dpac/consortium}). Funding for the DPAC has been provided by national institutions, in particular the institutions participating in the {\it Gaia} Multilateral Agreement.

\end{acknowledgements}

\bibliographystyle{aa-note} 

\end{document}